\documentclass[pra,twocolumn,aps,superscriptaddress,showpacs]{revtex4-2}
\usepackage{hyperref}
\hypersetup{
	colorlinks=true,
	citecolor=blue,
	linkcolor=blue,
	urlcolor=blue}
\usepackage{graphicx}
\usepackage{amsmath}
\usepackage{physics}
\usepackage{amsfonts}
\usepackage{amssymb}
\usepackage{bm}
\usepackage{siunitx}
\usepackage{xfrac}
\usepackage[normalem]{ulem}
\usepackage{braket}
\usepackage{cases}

\usepackage{graphicx}
\usepackage{psfrag}
\usepackage{dcolumn}
\usepackage{subfigure}
\usepackage{bm}
\usepackage{float}
\usepackage{xcolor}

\begin{document}

\title[]{Condensation temperature and magnetic phases of trapped coherently coupled Bose gases}

\author{Sunilkumar V}
\email{d21082@students.iitmandi.ac.in}
\affiliation{School of Physical Sciences, Indian Institute of Technology Mandi, Mandi-175075 (H.P.), India.}
\author{Rajat}%
 \email{rajat13sep@gmail.com}
 \affiliation{Department of Physics, Ben-Gurion University of the Negev, Beer-Sheva 84105, Israel}
\author{Sandeep Gautam}%
 \email{sandeep@iitrpr.ac.in}
 \affiliation{Department of Physics, Indian Institute of Technology Ropar, Rupnagar-140001, Punjab, India.}
 \author{Arko Roy}
 \email{arko@iitmandi.ac.in}
 \affiliation{School of Physical Sciences, Indian Institute of Technology Mandi, Mandi-175075 (H.P.), India.}

\begin{abstract}
 We investigate condensation and finite-temperature magnetic phase transitions in a 
 coherently (Rabi) coupled Bose gas confined in a three-dimensional harmonic trap. For the noninteracting system, we derive analytical expressions for the critical temperature and condensate fraction using a semiclassical description of the single-particle spectrum and density of states. At fixed particle number, coherent coupling enhances the critical temperature relative to the uncoupled system, and this enhancement decreases with increasing particle number $N$. The finite-size correction, in contrast, lowers the transition temperature, with its effect diminishing for larger $N$. We then incorporate repulsive interactions within the Hartree--Fock--Bogoliubov--Popov framework to investigate the finite-temperature phase diagram in the temperature--Rabi coupling plane. The system undergoes successive transitions from a ferromagnetic to a paramagnetic condensate and, at higher temperature, to a thermal gas. The transition boundaries obtained from the vanishing of the condensate fraction show excellent agreement with the analytical Hartree-Fock predictions. We find that the interactions substantially suppress the variation in the condensation temperature with an increase in the coherent coupling.
 \end{abstract}

\maketitle

\section{\label{intro}Introduction}

Recent advances in quantum simulation have opened new routes for 
realizing and probing dynamics in controllable quantum 
platforms, ranging from spin systems~\cite{Rutkevich_99,A.sinha_21,
Lagnese_24,johansen_25,luka_pavesic_25,daan_maertens_25,Borla_26} 
and digital quantum simulators~\cite{Abel_21,Vodeb_25} to ultracold 
atomic gases~\cite{Fialko_17,braden_18,Billam_20,Billam_21,Billam_22,
zenesini_24,Darbha_24,K.brown_25,jenkins_25}. Among these, 
coherently coupled Bose--Einstein condensates (BECs) are a particularly 
versatile platform to study analogue-gravity 
~\cite{PhysRevA.96.013611,fischer_04,garay_00,zenesini_24,
sivasankar2026},
quark-confinement~\cite{MinoruEto_18},
persistent currents~\cite{PhysRevA.93.033603}, internal Josephson
dynamics~\cite{T.Zibold_10,farolfi_2021_torque,Farolfi_21}, and to beyond-mean-field effects~\cite{Cappellaro2017,lavoine_21}
owing to their isolation from the environment~\cite{farolfi_19,rogora_24,Cominotti_2024}, 
tunable interactions~\cite{Bourdel_2022_tunable,sanz_22,Bourdel_2025}, 
and real-time access to nonequilibrium 
dynamics~\cite{farolfi_2021_torque,zenesini_24,cominotti_25}. 
At low temperatures, the two
hyperfine states form a spinor condensate coupled by an external 
electromagnetic (Rabi) field. The interplay between coherent coupling 
and unequal intra- and
interspecies interactions give rise to a ferromagnetic 
and a paramagnetic phase, separated by a quantum phase transition
characterized by spontaneous $\mathbb{Z}_2$ ordering of the longitudinal
spin component~\cite{oberthaler_2010,Abad2013}.  This transition can be mapped
onto the transverse-field Ising model, establishing coherently coupled
BECs as a quantum simulator of magnetic systems~\cite{recati_21,
cominotti_23}. Experimentally, such systems are realized in 
confining potentials, with harmonic traps providing the inhomogeneous 
density profiles characteristic of ultracold-atom experiments. This 
spatial inhomogeneity introduces an additional ingredient absent in 
homogeneous systems, as the local density and condensate properties vary across the trap.

The collective excitation spectrum of a coherently coupled BEC 
consists of a gapless density mode and a gapped spin mode, with the 
latter originating from the coherent coupling between the two 
components~\cite{recati_21,PhysRevA.55.2935,PhysRevA.64.013615,
PhysRevA.67.023606}. Thermal fluctuations are therefore expected to 
influence not only the condensate fraction but also the magnetic 
properties and collective excitations of the system. 
The finite-temperature phase diagram and collective modes of coherently coupled Bose mixtures have recently been 
investigated~\cite{Sunil_2026,PhysRevA.107.043301}. Furthermore, coherent 
coupling also provides a controlled route to nonequilibrium 
symmetry-breaking dynamics: universal scaling has been observed 
following quenches across critical 
points~\cite{PhysRevA.89.033631,Nicklas_2015}, and coupling quenches 
in binary condensates have been employed to study defect formation 
within the Kibble-Zurek mechanism~\cite{PhysRevLett.107.230402}. 
Temperature-driven false-vacuum decay has also been investigated in 
coherently coupled condensates~\cite{sivasankar2026}, where thermodynamic 
quantities such as the free energy provide a natural description of the 
decay process. A quantitative understanding of the thermodynamics, and 
in particular, of the critical temperature of the 
thermal-to-condensate transition, is therefore essential for understanding 
the behavior of coherently coupled gases upon cooling through their 
successive phases, from a thermal cloud to a paramagnetic condensate and 
ultimately to a ferromagnetic state. Such a cooling pathway provides 
a natural platform for exploring nonequilibrium phenomena, including 
Kibble--Zurek defect formation and subsequent coarsening dynamics.

The effect of coupling on the condensation temperature has been 
investigated in several related multicomponent Bose systems. 
For a homogeneous Rabi-coupled binary Bose gas, Ref.~\cite{Rakhimov2023}
obtained a positive shift of the critical temperature with 
increasing coherent coupling~\cite{Rakhimov2023}. The 
reported critical-temperature shift is determined by the one-body Rabi 
coupling, and the effect of interparticle interactions on $T_c$ is however, 
not captured~\cite{Rakhimov2023} within the Gaussian mean-field 
treatment.  Related studies of harmonically trapped spin-orbit-coupled Bose gases have
shown that coupling-induced modifications of the single-particle density of
states can significantly affect the critical temperature of an ideal gas with
finite particle number~\cite{Hui_hu_2012}, while self-consistent Hartree-Fock-Bogoliubov-Popov (HFB-Popov)
calculations find a Raman-coupling-dependent critical temperature from the
condensate fraction~\cite{PhysRevA.109.033319}. These studies establish 
two important aspects of the problem: coherent coupling can 
modify the condensation temperature, while the trapping geometry can 
qualitatively alter the role of such coupling through the density of 
states. However, the combined effect of coherent Rabi coupling, 
spatial inhomogeneity, and interparticle interactions on the condensation 
temperature remains unexplored.

In this work, we determine the critical temperature of a coherently 
Rabi-coupled Bose gas confined in a harmonic trap and investigate how 
it is modified by interparticle interactions. We first consider the 
noninteracting system, where coherent coupling modifies the single-particle 
spectrum and leads to an enhancement of the critical temperature. We then 
incorporate interactions and find that the increase of $T_c$ with coherent 
coupling is suppressed. Thus, our results connect the previously studied 
homogeneous Rabi-coupled system with the experimentally relevant 
inhomogeneous trapped geometry, while explicitly identifying the role of 
interactions in the coupling-induced shift of the condensation temperature.

The paper is organised as follows: Section \ref{ideal} develops the ideal-gas description, deriving the excitation spectrum and semiclassical density of states, followed by analytical results for the condensation temperature $T_c$ and condensate fraction of a coherently coupled Bose gas. Section \ref{finite} incorporates finite-size effects and obtains the corresponding corrections to $T_c$. Section \ref{interactions} introduces interactions within the HFB–Popov framework and derives the interaction-induced shift of $T_c$ employing the normal state HF theory. Section \ref{numerical} presents the self-consistent numerical results, establishing the finite-temperature phase diagram and the successive magnetic and condensation transitions, and compares the numerical transition temperatures with the analytical predictions. Section \ref{conclusions} summarizes our main findings.
\section{Non-interacting coherently coupled condensates in a harmonic trap}
\label{ideal}
\subsection{Excitation spectrum}
\label{excs}
We consider a coherently coupled non-interacting Bose–Bose mixture confined in a three-dimensional harmonic trap $V(\mathbf{r}) = m(\omega_x^2 x^2 + \omega_y^2 y^2 + \omega_z^2 z^2)/2$, where $\omega_x,\omega_y,\omega_z$ are the angular trapping frequencies along the $x,y,z$ directions. The single-particle Hamiltonian of the system is given by $H = \left[-({\hbar^2}/{2m})\nabla^2 + V(\mathbf r)\right]\mathbb{I} - \hbar\Omega\sigma_x$. The two components, each of having mass $m$ correspond to different hyperfine states $(\uparrow,\downarrow)$. The radio-frequency field with coupling strength $\hbar\Omega$ induces coherent transitions between these states, while $\mathbb{I}$ and $\sigma_x$ denote the identity and Pauli matrices, respectively. The coupling $\Omega$ is taken to be real and positive~\cite{stringari}.

To probe the excitation spectrum, we consider $\phi_n(\mathbf r)$ to be the eigenstate of the harmonic oscillator in the absence of coherent coupling with eigenenergies $\epsilon_n = \sum_{i=x,y,z}(n_i + 1/2)\hbar\omega_i$, where $n_i=0,1,2\cdots$ . We then construct the product state $\Phi_n(\mathbf r)=\phi_n(\mathbf r)\chi$, where $\chi$ is an eigenstate of $\sigma_x$. Operating with the Hamiltonian on $\Phi_n(\mathbf r)$ gives $H\Phi_n(\mathbf r) = (\epsilon_n\mathbb{I} - \hbar\Omega\sigma_x)\Phi_n(\mathbf r)$, from which the eigenenergies are readily obtained as $E_n^{\pm} = \epsilon_n \pm \hbar\Omega$. 
The coherent coupling splits each harmonic-oscillator level into two branches separated by $2\hbar\Omega$. The upper ($+$) and lower ($-$) branches are shifted linearly above and below the uncoupled spectrum, respectively. 
\begin{figure}
    \centering
\includegraphics[width=\columnwidth]{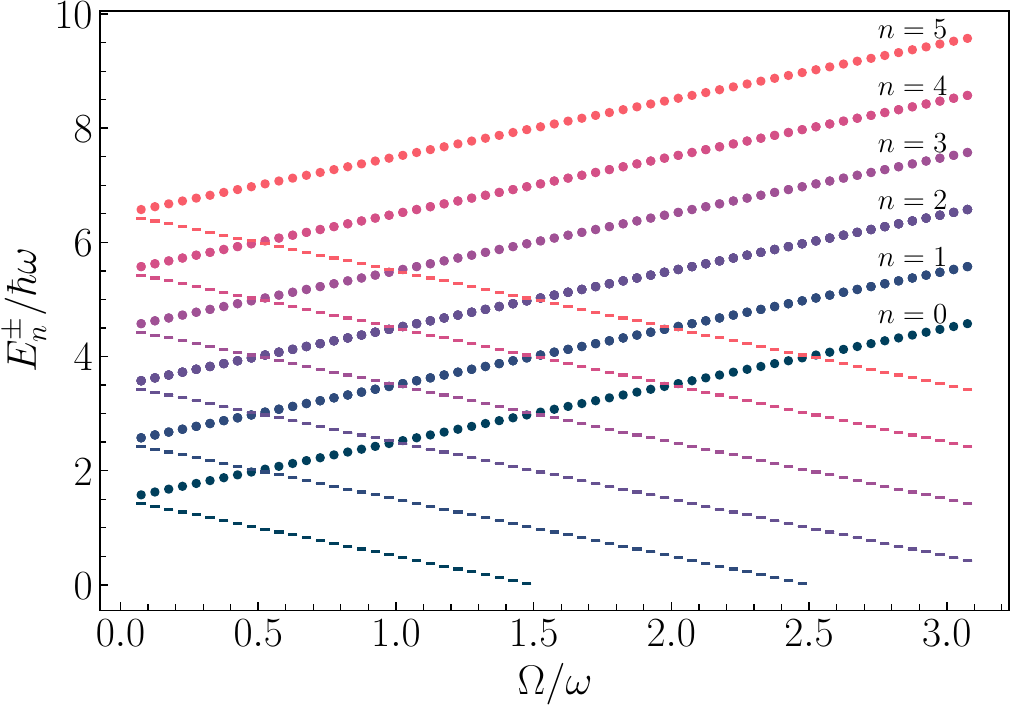}
    \caption{Excitation spectrum $E_n^{\pm}$ of an isotropic three-dimensional harmonically trapped, non-interacting coherently coupled BEC as a function of the coherent coupling strength. The dots represent the upper branch $E_n^{+}$, while the dashed lines denote the lower branch $E_n^{-}$.
    }
   \label{ex_spc}
\end{figure}
Fig.~\ref{ex_spc} illustrates this level splitting for an isotropic trap ($\omega_x=\omega_y=\omega_z=\omega$), where the excitation energies vary linearly with the dimensionless coupling strength $\Omega/\omega$.
\subsection{ Density of states in semiclassical approximation}
\label{dos}
The total number of particles $N$ in the grand-canonical ensemble at temperature $T$ is given by
\begin{equation}
N=\sum_{\alpha} \sum_{n}\frac{1}{e^{(E_n^{\alpha}-\mu)/k_BT}-1},
\label{N}
\end{equation}
where $\alpha=(+,-), n=(n_x,n_y,n_z)$,  and $\mu$ is the chemical potential. 

For large particle numbers relevant to the thermodynamic limit~\cite{stringari}, and when the excitation energies are much larger than the harmonic oscillator level spacing $\hbar\omega$, the discrete sum in Eq.~(\ref{N}) can be replaced by an integral, and zero-point energy can be neglected. This semiclassical approximation is valid in the regime $k_BT \gg \hbar\omega$~\cite{franco_1999}. Within this approximation, Eq.~(\ref{N}) becomes
\begin{equation}
N=\int dE\,\frac{\rho(E)}{e^{(E-\mu)/k_BT}-1},
\label{int_N}
\end{equation}
where $\rho(E)$ is the single-particle DOS.

For a harmonically trapped system, the DOS in 3D and 2D can be written as
\begin{eqnarray}
 \rho_{3D}(E)&=&\sum_{\pm}\sum_{n_x,n_y,n_z}\delta(E-E_n^{\pm}),\nonumber \\
 \rho_{2D}(E)&=&\sum_{\pm}\sum_{n_x,n_y}\delta(E-E_n^{\pm}).
 \label{dos32}
\end{eqnarray}
Under the semiclassical approximation, the summations over quantum numbers are replaced by integrals, yielding
\begin{equation}
\rho_{3D}(E)=\sum_{\pm}\iiint dn_x\,dn_y\,dn_z\,\delta(E-E_n^{\pm}).
\end{equation}
\begin{figure}
    \centering   \includegraphics[width=\columnwidth]{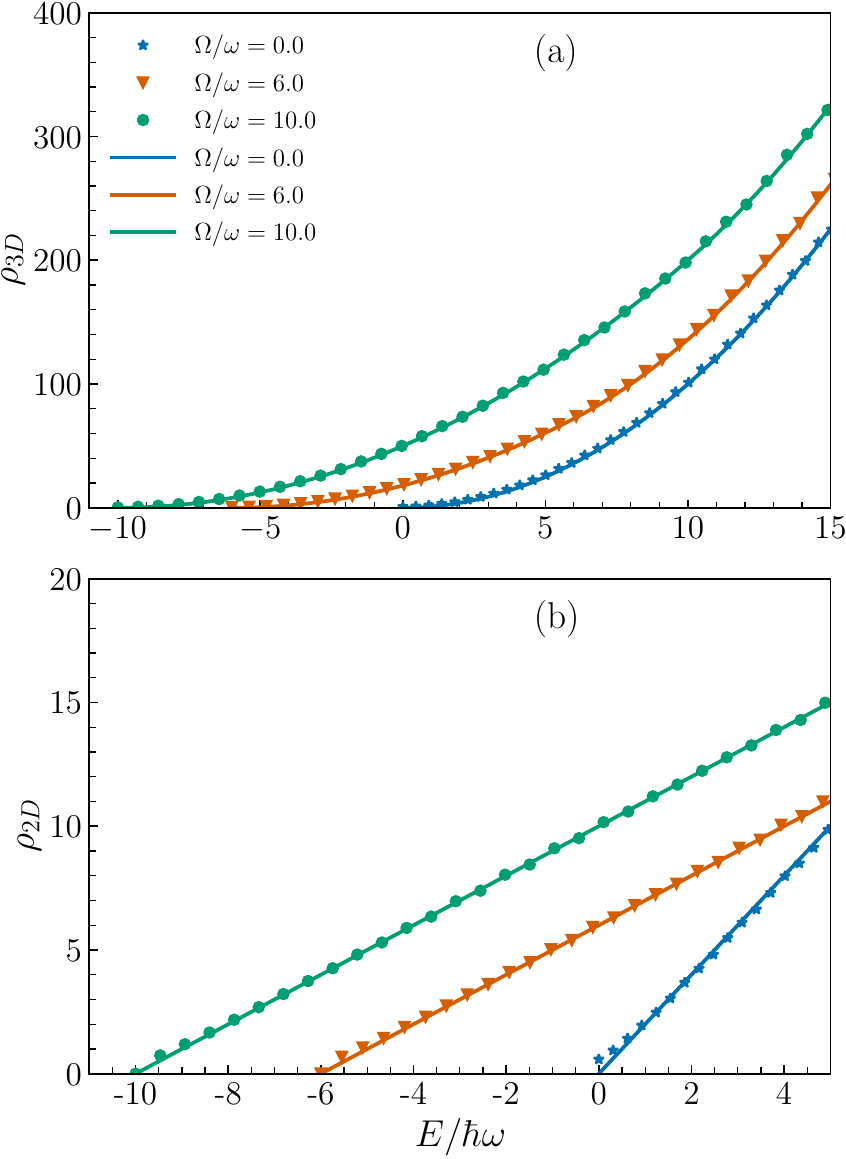}
    \caption{ Density of states (DOS) as a function of excitation energy for different coherent coupling strengths. (a) Three-dimensional DOS, $\rho_{3D}$, showing the expected quadratic dependence on energy. The solid lines correspond to the analytical semiclassical results, while the markers are obtained by directly evaluating Eq.~(\ref{dos32}) using the Lorentzian approximation to the Dirac delta function for an isotropic harmonic trap. (b) Two-dimensional DOS, $\rho_{2D}$, exhibiting the characteristic linear dependence on energy for different coherent coupling strengths. }
   \label{pd}
\end{figure}

Evaluating the integral yields the DOS in 3D (refer to Appendix~\ref{dos3d} for details),
\begin{equation}
\rho_{3D}(E)=\frac{\Theta(E+\Delta)}{2\hbar^3\omega_x\omega_y\omega_z}(E+\Delta)^2
+\frac{\Theta(E-\Delta)}{2\hbar^3\omega_x\omega_y\omega_z}(E-\Delta)^2 ,
\end{equation}
where $\Delta=\hbar\Omega$ and $\Theta(\cdot)$ denotes the Heaviside step function. 
Equivalently,
\begin{equation}
\rho_{3D}(E)=
\begin{cases}
0, & E<-\Delta,\\[4pt]
\dfrac{(E+\Delta)^2}{2\hbar^3\omega_x\omega_y\omega_z}, & -\Delta \le E < \Delta,\\[8pt]
\dfrac{(E+\Delta)^2+(E-\Delta)^2}{2\hbar^3\omega_x\omega_y\omega_z}, & E\ge \Delta .
\end{cases}
\label{dos_3D}
\end{equation}

In the absence of coherent coupling ($\Delta=0$), Eq.~(\ref{dos_3D}) reduces to the well-known DOS of a two-component Bose gas in a 3D harmonic trap, $\rho_{3D}(E)=E^2/(\hbar^3\omega_x\omega_y\omega_z)\Theta(E)$~\cite{Pethick_Smith_2001}.

Following the same procedure for the 2D excitation spectrum $E_n^{\pm}=\sum_{i=x,y}(n_i+\tfrac12)\hbar\omega_i\pm\hbar\Omega$, we obtain the DOS as
\begin{equation}
\rho_{2D}(E)=\frac{\Theta(E+\Delta)}{\hbar^2\omega_x\omega_y}(E+\Delta)
+\frac{\Theta(E-\Delta)}{\hbar^2\omega_x\omega_y}(E-\Delta).
\label{dos_2D}
\end{equation}
For $\Delta=0$, Eq.~(\ref{dos_2D}) reduces to the DOS of a two-component Bose gas in a 2D harmonic trap, $\rho_{2D}(E)=E/(\hbar^2\omega_x\omega_y)\Theta(E)$. 
The density of states in 3D and 2D for different coherent coupling strengths are shown in Fig.~\ref{pd}.

Equations~(\ref{dos_3D}) and (\ref{dos_2D}) show that the DOS increases quadratically with energy in 3D and linearly in 2D.  To validate the semiclassical expressions for DOS, we numerically evaluate Eq.~(\ref{dos32}).
In the numerical calculation, the Dirac delta function is approximated by the Lorentzian broadening function~\cite{Hui_hu_2012},
$f_{\delta}(x,\Gamma) = (1/\pi)\Gamma/(x^2 + \Gamma^2)$
with $x=E-E_n^\pm$ and $\Gamma\sim\hbar\omega$. For an isotropic trap ($\omega_x=\omega_y=\omega_z=\omega$), the DOS is evaluated on a discrete energy grid by summing the contributions from all eigenenergies $E_n^\pm$ for $n_i\leq30$. Each eigenstate contributes a Lorentzian peak centered at $E_n^\pm$, and the total DOS is obtained by superposing these contributions over the entire spectrum.
To further reduce the broadening introduced by the Lorentzian approximation, we employ the improved representation $\delta(x) = 2f_{\delta}(x,\Gamma=\hbar\omega) - f_{\delta}(x,\Gamma=2\hbar\omega)$
which significantly sharpens the spectral peaks. The resulting DOS is shown by the markers in Fig.~\ref{pd}. We find excellent agreement between the numerical results obtained from Eq.~(\ref{dos32}) and the analytical DOS derived within the semiclassical approximation.
 \subsection{Critical temperature and condensate fraction}
\label{crit_temp}
We now determine the critical temperature of a noninteracting coherently coupled BEC  using the excitation spectrum and the corresponding semiclassical density of states derived above. Within the semiclassical approximation, the total number of particles can be expressed as the sum of the condensate and thermal populations. Since the ground state becomes macroscopically occupied below the transition temperature, Eq.~(\ref{int_N}) can be written as $N = N_0 + N_{\mathrm{exc}}$, where $N_0$ is the condensate population and $N_{\mathrm{exc}}$ denotes the number of thermally excited particles. Explicitly,
\begin{equation}
N = N_0 + \int_{E_0^{\mathrm{sc}}}^{\infty} dE\,
\frac{\rho(E)}{e^{(E-\mu)/k_BT}-1},
\end{equation}
where $E_0^{\mathrm{sc}}=-\Delta$ is the minimum semiclassical energy.
At the critical temperature ($T=T_c$), the condensate population vanishes ($N_0\rightarrow0$), while the chemical potential approaches the minimum of the excitation spectrum, i.e., $\mu=E_0^{\mathrm{sc}}=-\Delta$. Consequently, the total particle number is entirely accounted for by the thermal excitations,
\begin{equation}
N = \int_{E_0^{\mathrm{sc}}}^{\infty} dE\,
\frac{\rho(E)}{e^{(E-\mu)/k_BT_c}-1}.
\label{exc}
\end{equation}
The critical temperature is obtained by evaluating the above integral using the appropriate semiclassical density of states.  

For the 3D case, introducing the variable $\chi=E+\Delta$ and substituting $\rho_{3D}(\chi)$ into Eq.~(\ref{exc}), we get 
\begin{equation}
N=\frac{1}{A}\left[\int_{0}^{2\Delta}\frac{\chi^2}{e^{\chi/k_BT_c}-1}d\chi
+\int_{2\Delta}^{\infty}\frac{\chi^2+(\chi-2\Delta)^2}{e^{\chi/k_BT_c}-1}d\chi\right],
\end{equation}
where $A=2\hbar^3\omega_x\omega_y\omega_z$. On evaluating the integrals we get
\begin{equation}
k_BT_c(\Omega)=
\frac{\hbar(N\omega_x\omega_y\omega_z)^{1/3}}
{\left[\zeta(3)+g_3\!\left(e^{-2\hbar\Omega/k_BT_c}\right)\right]^{1/3}},
\label{tc_3d}
\end{equation}
which determines the critical temperature of a non-interacting coherently coupled BEC in a 3D harmonic trap. Here $\zeta(3)=\displaystyle \sum_{n=1}^{\infty}1/n^3$ is the Riemann zeta function and $g_3(z)= \displaystyle \sum_{n=1}^{\infty}z^n/n^3$ is the Bose function of order $s=3$~\cite{ketterle_1996}.
Equation~(\ref{tc_3d}) is solved numerically to obtain $T_c(\Omega)$ for a given set of parameters to determine its dependence on the coherent coupling strength.
\begin{figure}[H]
    \centering
\includegraphics[width=\columnwidth]{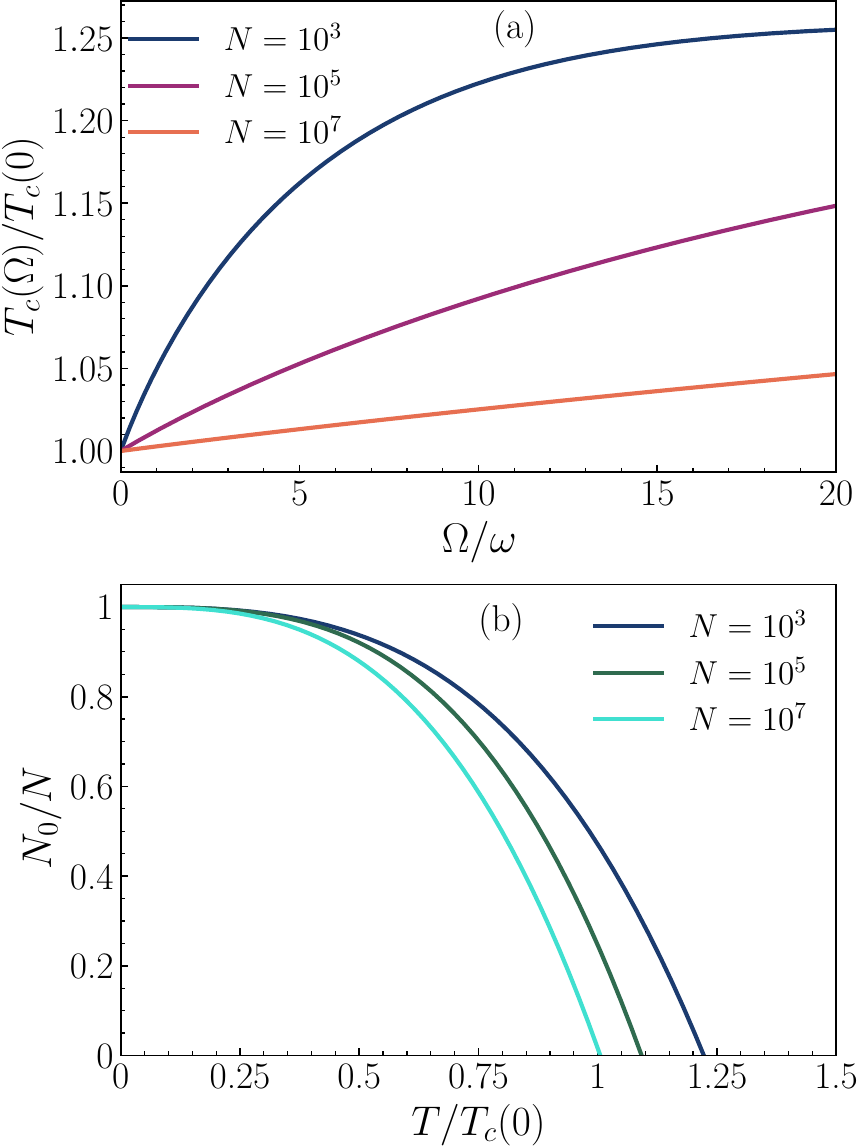}
    \caption{Variation of the critical temperature and condensate fraction of a three-dimensional ideal coherently coupled BEC. (a) Normalized critical temperature, $T_c(\Omega)/T_c(0)$, as a function of coherent coupling strength for different atom numbers. The enhancement of $T_c$ becomes less pronounced with increasing $N$. (b) Condensate fraction, $N_0/N$, as a function of the scaled temperature, $T/T_c(0)$, for different atom numbers at $\Omega/\omega=10$. For each atom number $N$, the temperature is normalized separately by its corresponding zero-coupling critical temperature. }
   \label{Tc_N_3d}
\end{figure}
For $\hbar\Omega=0$, one has $g_3(1)=\zeta(3)$, and Eq.~(\ref{tc_3d}) reduces to the well-known critical temperature of a two-component Bose gas in a 3D harmonic trap,
\[
k_B T_c(0)=\hbar\left[\frac{(N\omega_x\omega_y\omega_z)}{2\,\zeta(3)}\right]^{1/3},
\]
in agreement with Ref.~\cite{franco_1999}.
Figure~\ref{Tc_N_3d}(a) presents the variation of the normalized three-dimensional critical temperature, $T_c (\Omega)/T_c(0)$, with coherent coupling strength for different atom numbers. Note that in ultracold-atom experiments, harmonic trap frequencies are often tens to a few hundreds of Hz, whereas Rabi frequencies can readily be in the kHz range~\cite{lavoine_21}. In the strong-coupling limit, $\hbar\Omega \gg k_B T_c$, the upper excitation branch becomes thermally inaccessible, effectively freezing out excitations to $E_n^{+}$. Consequently, only the lower branch, $E_n^{-}$, contributes to the thermal population and the density of states. In this limit $g_3(e^{-2\hbar\Omega}/k_BT_c)\approx0$ reducing the critical temperature to $k_B T_c(\Omega) = \hbar\left((N\omega_x\omega_y\omega_z)/[\zeta(3)]\right)^{1/3}$, which is the critical temperature of a ideal single-component Bose gas in 3D harmonic trap.
 
To calculate condensate fraction for $T\le T_c(\Omega)$, we write, 
\begin{equation}
    \frac{N_0}{N} = 1- \frac{1}{N}\int_{E_{0}^{sc}}^{\infty} dE\frac{\rho(E)}{e^{(E-\mu)/k_BT}-1}.
\end{equation}
Solving the integral as before in evaluation of $T_c$, we get,

\begin{equation}
    \frac{N_0}{N} = 1 - \left(\frac{T}{T_c}\right)^3\frac{\left[\zeta(3) + g_3(e^{-2{\hbar\Omega}/{k_BT}})\right]}{\left[\zeta(3) + g_3(e^{-2{\hbar\Omega}/{k_BT_c}})\right]}.
    \label{frac_3d}
\end{equation}

Condensate fraction is plotted as a function of temperature $T$ for different number of atoms as shown in Fig.~\ref{Tc_N_3d}(b). 
Since $T_c(\Omega)>T_c(0)$, condensate fraction does not go to zero at $T/T_c(0) = 1.0$ but at some higher temperatures when  $T/T_c(0)= T_c(\Omega)/T_c(0)$, where $T_c(\Omega)$ is given by Eq.~(\ref{tc_3d}). Further, for $N\rightarrow \infty$, $T_c(\Omega)\rightarrow T_c(0)$ and hence condensate fraction becomes zero exactly at $T/T_c(0)=1.0$, consistent with Fig.~\ref{Tc_N_3d}(a).

For the two-dimensional case, the calculation proceeds in a similar manner to the 3D case. Using the corresponding semiclassical density of states, we obtain the critical temperature
\begin{equation}
k_BT_c(\Omega)=
\frac{\hbar(N\omega_x\omega_y)^{1/2}}
{\left[\zeta(2)+g_2\!\left(e^{-2\hbar\Omega/k_BT_c}\right)\right]^{1/2}},
\label{tc_2d}
\end{equation}
and the corresponding condensate fraction is
\begin{equation}
\frac{N_0}{N}
=
1-
\left(\frac{T}{T_c}\right)^2
\frac{\zeta(2)+
g_2\!\left(e^{-2\hbar\Omega/k_BT}\right)}
{\zeta(2)+
g_2\!\left(e^{-2\hbar\Omega/k_BT_c}\right)}.
\label{conf2d}
\end{equation}
where $\zeta(2)$ is the Riemann zeta function and $g_2(z)$ denotes the Bose function of order two. In the absence of coherent coupling, $g_2(1)=\zeta(2)$, and Eqs.~(\ref{tc_2d}),(\ref{conf2d}) indeed reduce to their respective values for an ideal two-component Bose gas in a two-dimensional harmonic trap.

\section{Finite-size correction to critical temperature}
\label{finite}

The analytical results obtained in the previous section are based on the semiclassical approximation, which neglects the zero-point energy of the harmonic trap and takes the minimum single-particle energy to be $E_0^{\mathrm{sc}}=-\Delta$. For a finite trapped system, however, the discrete nature of the trap spectrum leads to a finite zero-point energy and consequently modifies the minimum single-particle energy~\cite{grossman_1995,georgini_1996}. This results in a finite-size correction to the critical temperature. We therefore evaluate the critical temperature, including this correction, for three-dimensional harmonic traps, considering finite atom numbers and experimentally relevant trap frequencies. 
For a finite system, the exact ground-state energy is
\begin{equation}
E_0=\frac{\hbar}{2}(\omega_x+\omega_y+\omega_z)-\Delta,
\end{equation}
which differs from the semiclassical value by the zero-point energy. Since the chemical potential equals the lowest single-particle energy at the transition, the particle number is given by
\begin{equation}
N=\int_{E_0}^{\infty}
\frac{\rho_{3D}(E)}
{e^{(E-E_0)/k_BT_c}-1}\,dE.
\end{equation}
Introducing the shifted energy variable $E'=E-E_0$, the above expression becomes
\begin{equation}
N=
\int_0^{\infty}
\frac{\rho_{3D}(E'+E_0^{\mathrm{sc}}+\Delta E)}
{e^{E'/k_BT_c}-1}\,dE',
\label{finite_N}
\end{equation}
where $\Delta E=E_0-E_0^{\mathrm{sc}}>0$.
Equation~(\ref{finite_N}) shows that the finite-size correction is equivalent to evaluating the semiclassical density of states at energies shifted by the zero-point energy $\Delta E$. Consequently,  the excitation spectrum is displaced to higher energies compared to the semiclassical result of Eq.~(\ref{exc}), giving rise to the finite-size correction to the critical temperature.

\begin{figure}
    \centering
\includegraphics[width=\columnwidth]{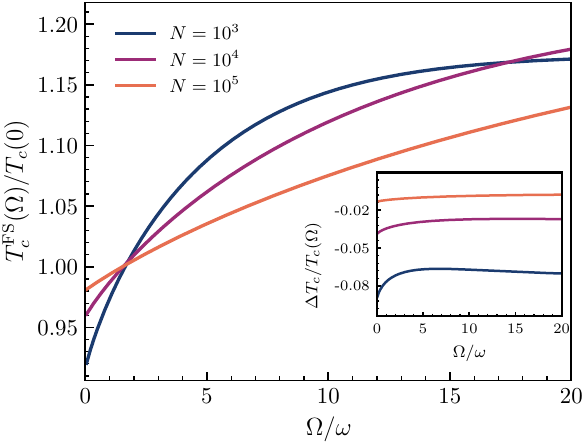}
    \caption{Finite-size correction to the critical temperature in 3D as a function of coherent coupling strength for different atom numbers. For each $N$, the temperature is normalized by its corresponding zero-coupling critical temperature. The inset shows the relative finite-size correction, $\Delta T_c/T_c(\Omega) = [T^{\rm FS}_c(\Omega)-T_c(\Omega)]/T_c(\Omega)$, as a function of the coherent coupling strength.}
    \label{finite_3d}
\end{figure}

For an isotropic trap, the zero-point energy correction satisfies $\Delta E\sim\hbar\omega$ and is the smallest energy scale. Therefore, the density of states can be expanded to first order in $\Delta E$ as
\[
\rho_{3D}(E'+E_0^{\mathrm{sc}}+\Delta E)
\simeq
\rho_{3D}(E'+E_0^{\mathrm{sc}})
+
\Delta E
\left.
\frac{\partial\rho_{3D}}{\partial E}
\right|_{E'+E_0^{\mathrm{sc}}}.
\]
Furthermore, the three- and two-dimensional densities of states satisfy the relation $\hbar\omega_z\,\partial\rho_{3D}(E)/\partial E=\rho_{2D}(E)$. Using these results, Eq.~(\ref{finite_N}) can be written as

\begin{equation}
N=
\int_{0}^{\infty}
\frac{\rho_{3D}(E'+E_0^{\mathrm{sc}})}
{e^{E'/k_BT_c}-1}\,dE'
+
\frac{\Delta E}{\hbar\omega_z}
\int_{0}^{\infty}
\frac{\rho_{2D}(E'+E_0^{\mathrm{sc}})}
{e^{E'/k_BT_c}-1}\,dE'.
\end{equation}
Solving the above integrals for $T_c$, we get the finite-size correction to the 3D critical temperature,
\begin{align}
N &= \frac{(k_{B}T_c)^3}{\hbar^3\omega_x\omega_y\omega_z}
\Bigg\{
\left[\zeta(3) + g_3\!\left(e^{-2\hbar\Omega/k_BT_c}\right)\right]
\nonumber\\
&\qquad
+ \frac{\Delta E}{k_BT_c}
\left[\zeta(2) + g_2\!\left(e^{-2\hbar\Omega/k_BT_c}\right)\right]
\Bigg\}.
\label{finite_correct_3d}
\end{align}
The first term in the above equation is identical to Eq.~(\ref{tc_3d}) and represents the thermodynamic-limit contribution, whereas the second term arises solely from the finite-size correction. Figure~\ref{finite_3d} compares the finite-size corrected critical temperature with the corresponding thermodynamic-limit result for different atom numbers. The finite-size effect is most pronounced for small systems, leading to a reduction in the critical temperature.  As the atom number increases, the correction gradually diminishes, and the finite-size result approaches the thermodynamic limit.
In the absence of coherent coupling ($\hbar\Omega=0$), Eq.~(\ref{finite_correct_3d}) yields the well-known finite-size correction for an ideal two-component Bose gas confined in a three-dimensional harmonic trap~\cite{Hui_hu_2012,ketterle_1996,HAUGERUD199718,georgini_1996}.

To obtain the finite-size correction to the critical temperature in two dimensions, we follow the same procedure as in the three-dimensional case in which 
the number equation at the critical temperature is obtained as
\begin{eqnarray}
N &=& \frac{(k_{B}T_c)^2}{\hbar^2\omega_x\omega_y}
\bigg\{
\left[\zeta(2)+g_2(e^{-2\hbar\Omega/k_BT_c})\right]
\nonumber\\
&+&\frac{\Delta E}{k_BT_c}
\left[
\ln\!\bigg(\frac{2k_BT_c}{\hbar\omega_x}\right)\nonumber\\
&-&\ln\!\left(1-e^{-(2\Delta+\hbar\omega_x/2)/k_BT_c}\bigg)
\right]
\bigg\}.
\end{eqnarray}
As in the three-dimensional case, the finite-size correction lowers the critical temperature, with its magnitude decreasing progressively as the number of atoms increases. Finite-size corrections to the critical temperature for the 1D system are discussed in Appendix~\ref{fs1d}.
\section{Interacting coherently coupled gas and correction to $T_c$}
\label{interactions}
\subsection{Semiclassical HFB-Popov equations}
We now develop the finite-temperature description of the interacting gas and
then apply it to the condensation boundary.
Since interactions significantly influence the thermodynamic properties of dilute Bose gases, they are expected to modify the transition temperature. We begin with the Heisenberg equation of motion for the Bose field operators

\begin{eqnarray}
    i \hbar\begin{pmatrix}
  \hat{\dot\Psi}_\uparrow\\
  \hat{\dot\Psi}_\downarrow 
  \end{pmatrix}
  &=&
  \begin{pmatrix}
      \hat{h} + g_{\uparrow \uparrow}\hat\Psi^{\dag}_{\uparrow}\hat\Psi_{\uparrow} & g_{\uparrow \downarrow}\hat\Psi^{\dag}_{\downarrow}\hat\Psi_{\uparrow} \\
      g_{\uparrow \downarrow}\hat\Psi^{\dag}_{\uparrow}\hat\Psi_{\downarrow} & \hat{h} + g_{\downarrow \downarrow}\hat\Psi^{\dag}_{\downarrow}\hat\Psi_{\downarrow}
  \end{pmatrix}
  \begin{pmatrix}
  \hat\Psi_\uparrow\\
  \hat\Psi_\downarrow  
\end{pmatrix}
  \nonumber\\
  &-&
  \begin{pmatrix}
        0 & \hbar\Omega\\
        \hbar\Omega & 0
     \end{pmatrix}
     \begin{pmatrix}
  \hat\Psi_\uparrow\\
  \hat\Psi_\downarrow  
\end{pmatrix},
 \label{heisenberg}
\end{eqnarray}
where
$\hat{h}=-{(\hbar^2\nabla^2)}/{2m}+V(\mathbf{r})-\mu$. The interaction strengths $g_{\uparrow\uparrow}$ and $g_{\downarrow\downarrow}$ denote the intraspecies interactions, while $g_{\uparrow\downarrow}$ characterizes the interspecies interaction.
To describe the interacting condensate at finite temperature, we employ the HFB theory within the Popov approximation (HFB-Popov)~\cite{Sunil_2026}. The Bose field operator is decomposed as
$\hat{\Psi}_i(\mathbf{r},t)=\phi_i(\mathbf{r})+\delta\hat{\psi}_i(\mathbf{r},t)$,
where $\phi_i$ is the condensate wavefunction and $\delta\hat{\psi}_i$ represents thermal fluctuations. Assuming
$
\langle\delta\hat{\psi}_i\rangle
=
\langle\delta\hat{\psi}_i^\dagger\rangle
=
0,
$
and neglecting the anomalous averages,
$
\langle\delta\hat{\psi}_i\delta\hat{\psi}_i\rangle,
\qquad
\langle\delta\hat{\psi}_i\delta\hat{\psi}_{\bar{i}}\rangle,
$
the generalized Gross-Pitaevskii equations reduce to

\begin{equation} 0=
\begin{pmatrix}
 \hat{h} + L_{\uparrow} & g_{\uparrow\downarrow}\langle\delta{\hat\psi}_{\uparrow}^\dag \delta{\hat\psi}_{\downarrow}\rangle -\hbar\Omega\\
 g_{\uparrow\downarrow}\langle\delta{\hat\psi}_{\uparrow}^\dag \delta{\hat\psi}_{\downarrow}\rangle -\hbar\Omega & \hat{h} + L_{\downarrow}
\end{pmatrix}
\begin{pmatrix}
    \phi_{\uparrow}\\\phi_{\downarrow}
\end{pmatrix},
\label{order1}
\end{equation}
where
$L_i=g_{ii}(n_i+\tilde n_i)+g_{i\bar{i}}n_{\bar{i}}$,
$\tilde n_i=\langle\delta\hat{\psi}_i^\dagger\delta\hat{\psi}_i\rangle$ is the non-condensate density,
$n_i=|\phi_i|^2+\tilde n_i$
is the total density of the $i$th component and $\tilde {n}_{i\bar{i}} =  \langle\delta{\hat\psi}_{i}^\dag \delta{\hat\psi}_{\bar{i}}\rangle$ is the coherence between thermal atoms. Here $\bar i = \downarrow (\uparrow)$ if $ i = \uparrow (\downarrow)$.
Subtracting the generalized Gross-Pitaevskii equations [Eq.~(\ref{order1})] from the Heisenberg equations of motion [Eq.~(\ref{heisenberg})] within the mean-field approximation~\cite{Proukakis_2008}, yields the equations of motion for the fluctuation operators, 
\begin{eqnarray}
   i\hbar\frac{\partial\delta{\hat{\psi}_\uparrow}}{\partial t} &= &\left[\hat{h} + 2g_{\uparrow \uparrow}(n_{\uparrow c} + {\tilde{n}_\uparrow}) + g_{\uparrow\downarrow} (n_{\downarrow c} + {\tilde{n}_\downarrow})\right]\delta{\hat{\psi}_\uparrow}
   \nonumber\\
   &+& g_{\uparrow \uparrow}\phi_{\uparrow}^2 \delta \hat{\psi}_{\uparrow}^{\dag} + [g_{\uparrow\downarrow}(\phi_{\uparrow}\phi_{\downarrow}^{*} + \tilde n_{\downarrow \uparrow}) -\hbar\Omega]\delta{\hat{\psi}_\downarrow}\nonumber\\
   &+& g_{\uparrow\downarrow}\phi_\uparrow \phi_\downarrow \delta{\hat{\psi}_{\downarrow}}^{\dag},  \nonumber\\
   i\hbar\frac{\partial\delta{\hat{\psi}_\downarrow}}{\partial t} &=& \left[\hat{h} + 2g_{\downarrow\downarrow}(n_{\downarrow c} + {\tilde{n}_\downarrow}) + g_{\uparrow\downarrow} (n_{\uparrow c} + {\tilde{n}_\uparrow})\right]\delta{\hat{\psi}_\downarrow} \nonumber\\
   &+& g_{\downarrow\downarrow}\phi_{\downarrow}^2 \delta\hat{\psi}_{\downarrow}^{\dag} \nonumber+
  [g_{\uparrow\downarrow}(\phi_{\downarrow}\phi_{\uparrow}^{*} + \tilde n_{\uparrow \downarrow}) - \hbar\Omega]\delta{\hat{\psi}_\uparrow} \nonumber\\
  &+& g_{\uparrow\downarrow}\phi_\downarrow \phi_\uparrow \delta{\hat{\psi}_{\uparrow}}^{\dag}.
   \label{fluct1}
\end{eqnarray}

\subsection{Bogoliubov spectrum and spectral weights}
Starting from the above equations, we formulate the Bogoliubov--de Gennes (BdG) problem in the bare-component basis
$W_\alpha=
\left(
u_{\uparrow\alpha},
u_{\downarrow\alpha},
v_{\uparrow\alpha},
v_{\downarrow\alpha}
\right)^T,$
where
$\alpha=-,+$ labels the two positive-energy BdG branches,
and $u_i$ and $v_i$ denote the quasiparticle amplitudes of the two components~\cite{Sunil_2026}. 
In the semiclassical approximation the
kinetic-energy operator is replaced by
$-\hbar^2\nabla^2/(2m)\rightarrow p^2/(2m)$. The resulting BdG equations can be written in the generalized eigenvalue form
\begin{equation}
\mathcal M W_\alpha
=\epsilon_\alpha\sigma W_\alpha,
\quad
W_\alpha^\dagger\sigma W_\beta=\delta_{\alpha\beta},
\label{generalized_bdg_problem}
\end{equation}
where $\mathcal M=\mathcal M^\dagger$,
$\epsilon$ is the quasiparticle energy and $\sigma=\mathrm{diag}(1,1,-1,-1)$.  Considering equal repulsive intraspecies interaction strengths, $g_{\uparrow\uparrow}=g_{\downarrow\downarrow}=g$, the BdG matrix is given by

\begin{equation}
\mathcal{M}=
\begin{pmatrix}
H_{d,\uparrow} & H_{od}
& \Delta_{\uparrow\uparrow} & \Delta_{\uparrow\downarrow}
\\
H_{od} & H_{d,\downarrow}
& \Delta_{\uparrow\downarrow} & \Delta_{\downarrow\downarrow}
\\
\Delta_{\uparrow\uparrow} & \Delta_{\uparrow\downarrow}
& H_{d,\uparrow} & H_{od}
\\
\Delta_{\uparrow\downarrow} & \Delta_{\downarrow\downarrow}
& H_{od} & H_{d,\downarrow}
\end{pmatrix},
\label{bdg_matrix}
\end{equation}
with the matrix elements 

\begin{eqnarray}
H_{d,\uparrow} &=& \frac{p^2}{2m} + V(\mathbf{r}) - \mu + 2g n_\uparrow + g_{\uparrow\downarrow} n_\downarrow \nonumber\\
H_{d,\downarrow} &=& \frac{p^2}{2m} + V(\mathbf{r}) - \mu + 2g n_\downarrow + g_{\uparrow\downarrow} n_\uparrow\nonumber\\
H_{od} &=& g_{\uparrow\downarrow} (\sqrt{n_{c\uparrow}n_{c\downarrow}} +  \tilde{n}_{\uparrow\downarrow})- \hbar\Omega \hspace{1.3cm}\nonumber \\
\Delta_{\uparrow\uparrow} &=& g n_{c\uparrow}, \quad \Delta_{\downarrow\downarrow} = g n_{c\downarrow} \nonumber \hspace{2cm}\\
\Delta_{\uparrow\downarrow} &=& g_{\uparrow\downarrow} \sqrt{n_{c\uparrow} n_{c\downarrow}}. \hspace{3.1cm}
\end{eqnarray}

The excitation energies are given by,
\begin{equation}
\begin{aligned}
\epsilon_{\pm}
&=
\left[
\frac{\Lambda\pm\sqrt{\Lambda^2-4\Gamma}}{2}
\right]^{1/2},
\end{aligned}
\label{eigen_value}
\end{equation}

where
\begin{equation}
    \Lambda = H_{d,\uparrow}^2 + H_{d,\downarrow}^2 + 2H_{od}^2 - \Delta_{\uparrow\uparrow}^2 - \Delta_{\downarrow\downarrow}^2 - 2\Delta_{\uparrow\downarrow}^2,\nonumber
    \end{equation}
and
\begin{equation}
    \begin{split}
        \Gamma &= \left[ \left(H_{d,\uparrow} - \Delta_{\uparrow\uparrow}\right) \left(H_{d,\downarrow} - \Delta_{\downarrow\downarrow}\right) - \left(H_{od} - \Delta_{\uparrow\downarrow}\right)^2 \right] \\
        &\quad \times \left[ \left(H_{d,\uparrow} + \Delta_{\uparrow\uparrow}\right) \left(H_{d,\downarrow} + \Delta_{\downarrow\downarrow}\right) - \left(H_{od} + \Delta_{\uparrow\downarrow}\right)^2 \right].\nonumber
    \end{split}
\end{equation}

Here, the subscripts $d$ and $od$ denote the diagonal and off-diagonal components in the bare-component representation, respectively. 
Eq.~(\ref{eigen_value}) gives the two positive-energy branches of
the coherently coupled two-component BdG spectrum. The thermal densities must therefore be resolved into the two bare components using the corresponding component-resolved spectral weights.
For completeness, we derive here the spectral weight identity used below. Differentiating
Eq.~(\ref{generalized_bdg_problem}) with respect to a matrix element
$X$ gives
\begin{equation}
(\partial_X\mathcal M)W_\alpha
+\mathcal M(\partial_XW_\alpha)
=
(\partial_X\epsilon_\alpha)\sigma W_\alpha
+\epsilon_\alpha\sigma(\partial_XW_\alpha).
\label{diff}
\end{equation}
The Hermitian conjugate of eigenvalue Eq.~(\ref{generalized_bdg_problem}) is
\begin{equation}
W_\alpha^\dagger\mathcal M
=\epsilon_\alpha W_\alpha^\dagger\sigma.
\end{equation}
Multiplying Eq.~(\ref{diff}) from the left by $W_\alpha^\dagger$, yields
\begin{equation}
\frac{\partial\epsilon_\alpha}{\partial X}
=
\frac{
W_\alpha^\dagger(\partial_X\mathcal M)W_\alpha
}{
W_\alpha^\dagger\sigma W_\alpha
},
\end{equation}
and with the positive-norm convention in Eq.~(\ref{generalized_bdg_problem}), this reduces to
\begin{equation}
\frac{\partial\epsilon_\alpha}{\partial X}
=
W_\alpha^\dagger
\frac{\partial\mathcal M}{\partial X}
W_\alpha.
\label{generalized_hf_identity}
\end{equation}
Applying Eq.~(\ref{generalized_hf_identity}) to the diagonal and
off-diagonal matrix elements gives the component-resolved spectral
weights
\begin{subequations}
\begin{align}
w_{i,\alpha}&=\frac{\partial\epsilon_\alpha}{\partial H_{d,i}}
=|u_{i\alpha}|^2+|v_{i\alpha}|^2,
\label{spectral_weight}
\\
w_{\uparrow\downarrow,\alpha}
&=\frac12\frac{\partial\epsilon_\alpha}{\partial H_{od}}
=\operatorname{Re}\!\left(u_{\downarrow\alpha}^{*}u_{\uparrow\alpha}
+v_{\downarrow\alpha}v_{\uparrow\alpha}^{*}\right),
\label{spectral_weight_12}
\end{align}
\end{subequations}
 Here $w_{i,\alpha}$ is the weight of bare component $i$ in BdG branch
$\alpha$, while $w_{\uparrow\downarrow,\alpha}$ is the corresponding
weight of coherence terms.

Since the chemical potential enters both diagonal elements as $-\mu$, these weights obey the sum rule
\begin{equation}
-\left(\frac{\partial\epsilon_\alpha}{\partial\mu}\right)_{\rm mf}
=w_{\uparrow,\alpha}+w_{\downarrow,\alpha},
\label{spectral_weight_sum}
\end{equation}
where the subscript ``mf'' means that the condensate and thermal mean fields
are held fixed.
Based on these considerations, the thermal part of the
non-condensate density and thermal coherence are given by
\begin{subequations}
\begin{align}
\tilde n_i(\mathbf r)&=\int\frac{d^3p}{(2\pi\hbar)^3}
\sum_{\alpha=\pm}w_{i,\alpha}f(\epsilon_\alpha),
\label{ntilde_bdg}\\
\tilde n_{\uparrow\downarrow}(\mathbf r)&=
\int\frac{d^3p}{(2\pi\hbar)^3}
\sum_{\alpha=\pm}w_{\uparrow\downarrow,\alpha}f(\epsilon_\alpha).
\label{ntilde12_bdg}
\end{align}
\end{subequations}
where $f(\epsilon_\alpha)=[\exp(\beta\epsilon_\alpha)-1]^{-1}$ is the Bose-Einstein distribution function. Quantum-depletion terms are neglected in the semiclassical HFB-Popov approximation~\cite {georgini_1996}.

\subsection{Normal-state Hartree-Fock theory}
We will now determine the interaction-induced shift of the condensation temperature $T_c$.
 At the condensation
boundary $\Delta_{\uparrow\uparrow}= \Delta_{\downarrow\downarrow}=\Delta_{\uparrow\downarrow}=0$ since $n_{\uparrow c}(\mathbf{r})=n_{\downarrow c}(\mathbf{r})=0$.
Additionally, since $v_{\uparrow} = v_{\downarrow}=0$, the problem reduces to the diagonalization of the following
$2\times2$ Hartree-Fock (HF) matrix,
\begin{equation}
\mathcal M_{\rm HF}=
\begin{pmatrix}
H_{d,\uparrow} & H_{od}\\
H_{od} & H_{d, \downarrow}
\end{pmatrix},
\end{equation}
where
\begin{equation}
 H_{d,i}=\frac{p^2}{2m}+V(\mathbf r)-\mu
+2g \tilde n_{i}+g_{\uparrow\downarrow}\tilde n_{\bar i},\ H_{od} =g_{\uparrow \downarrow }\tilde n_{\uparrow \downarrow}-\hbar\Omega.
\end{equation}
The eigenvalues are
\begin{equation}
\epsilon_\pm^{\rm HF}=\frac{H_{d,\uparrow}+ H_{d,\downarrow}}{2}
\pm\sqrt{\left(\frac{H_{d,\uparrow}-H_{d,\downarrow}}{2}\right)^2
+H_{od}^2}.
\label{general_hf_spectrum}
\end{equation}
The component weights follow directly from the
derivatives of the two HF eigenvalues
\begin{align}
w^{\rm HF}_{\uparrow,\pm}
&=\frac{\partial\epsilon_\pm^{\rm HF}}{\partial H_{d,\uparrow}}
=\frac12\pm\frac{H_{d,\uparrow}-H_{d,\downarrow}}
{4\sqrt{[(H_{d,\uparrow}- H_{d,\downarrow})/2]^2
+H_{od}^2}},\nonumber\\
w^{\rm HF}_{\downarrow,\pm}
&=\frac{\partial\epsilon_\pm^{\rm HF}}{\partial H_{d,\downarrow}}
=1-w^{\rm HF}_{\uparrow,\pm},\nonumber\\
w^{\rm HF}_{\uparrow\downarrow,\pm}
&=\frac12\frac{\partial\epsilon_\pm^{\rm HF}}{\partial H_{od}}
=\pm\frac{H_{od}}{2\sqrt{[(H_{d,\uparrow}- H_{d,\downarrow})/2]^2
+H_{od}^2}}.
\label{hf_component_weights}
\end{align}
The thermal densities and coherence terms are 
\begin{subequations}
\begin{align}
\tilde n_i(\mathbf r)&=\int\frac{d^3p}{(2\pi\hbar)^3}
\sum_{\alpha=\pm}w^{\rm HF}_{i,\alpha}f(\epsilon_\alpha^{\rm HF}),
\label{normal_hf_densities}\\
\tilde n_{\uparrow\downarrow}(\mathbf r)&=
\int\frac{d^3p}{(2\pi\hbar)^3}
\sum_{\alpha=\pm}w^{\rm HF}_{\uparrow\downarrow,\alpha}f(\epsilon_\alpha^{\rm HF}),
\label{normal_hf_coherence}
\end{align}
\end{subequations}
where $f(\epsilon_\alpha^{\rm HF})$ is the usual BE distribution function.

For an energetically favorable balanced normal coherently coupled gas $\tilde n_\uparrow=\tilde n_\downarrow=\tilde n/2$, the two HF eigenvalues given in Eq.~(\ref{general_hf_spectrum}) take the form
\begin{equation}
\epsilon_{\pm}^{\rm HF} = \frac{p^2}{2m}+V-\mu + C\tilde n
\pm (\hbar\Omega-g_{\uparrow\downarrow}\tilde n_{\uparrow\downarrow}),
\end{equation}
where $C=g+g_{\uparrow\downarrow}/2$. Here the minus and plus signs
label the lower and upper HF branches, respectively.
The spectral weights from Eq.~(\ref{hf_component_weights}) reduce to $w^{\rm HF}_{i,\pm}=1/2$, $w^{\rm HF}_{\uparrow\downarrow,-}=-w^{\rm HF}_{\uparrow\downarrow,+}=1/2$.
 
Introducing the densities of the lower and upper HF branches,
\begin{equation}
\tilde n_\pm(\mathbf r)
=\frac{1}{\lambda_T^3}
g_{3/2}\!\left[z_\pm(\mathbf r)\right],
\end{equation}
where $z_{\pm }({\bf r}) = \exp[-\beta\{\epsilon_{\pm}^{\rm HF}({\bf r}, {\bf p}) -p^2/(2m)\}]$,
$
\lambda_T=\hbar\sqrt{{2\pi}/{mk_BT}}
$
is the thermal de Broglie wavelength and
$
g_{3/2}(x)=\sum_{l=1}^{\infty}{x^l}/{l^{3/2}}
$
is the Bose function, and following Eqns.~(\ref{normal_hf_densities}) and (\ref{normal_hf_coherence}), the
semiclassical bare-component HF thermal densities and thermal coherence are given by
\begin{subequations}
\begin{align}
\tilde n_\uparrow(\mathbf r)
=&\tilde n_\downarrow(\mathbf r)
=\frac{\tilde n_-(\mathbf r)+\tilde n_+(\mathbf r)}{2},\\
\tilde n_{\uparrow\downarrow}(\mathbf r)
=&\frac{\tilde n_-(\mathbf r)-\tilde n_+(\mathbf r)}{2}.
\label{balanced_hf_coherence}
\end{align}
\end{subequations}

\subsection{First-order interaction shift of the critical temperature}
A closed analytical result can be obtained by expanding the HF thermal densities to first order in the
interactions. To this end, we introduce the ideal-gas thermal densities associated with
the upper (+) and lower (-) single-particle branches,
\begin{align}
\tilde n_{\pm}^0({\bf r}) =\frac{1}{\lambda_T^{3}}
g_{3/2}
\!\left[z_{\pm}^0({\bf r})\right], \label{n_T_non-int}
\end{align}
where $z_{\pm}^0 = z_{\pm}\vert_{g=0,g_{\uparrow\downarrow} =0}$.
Defining $\tilde n^0({\bf r})=\tilde n_{+}^0({\bf r})+\tilde n_{-}^0({\bf r})$, and $\tilde n_{\uparrow \downarrow}^0({\bf r})=(\tilde n_{-}^0({\bf r})-\tilde n_{+}^0({\bf r}))/2$, 
and expanding the two branch densities, $\tilde{n}_{\pm}({\bf r})$, to first order in the interactions about $z_{\pm}^0$ (assuming $\beta (C\tilde n
\mp g_{\uparrow\downarrow}\tilde n_{\uparrow\downarrow})\ll 1$) 
gives the thermal density $n_T({\bf r})$ ~\cite{georgini_1996},
\begin{equation}
n_T({\bf r})\simeq\tilde n^0({\bf r})
-C\tilde n^0({\bf r})
\frac{\partial\tilde n^0({\bf r})}{\partial\mu}+g_{\uparrow \downarrow}\tilde n_{\uparrow \downarrow}^0({\bf r})
\left[\frac{\partial\tilde n_{+}^0({\bf r})}{\partial\mu}
-\frac{\partial\tilde n_{-}^0({\bf r})}{\partial\mu}\right].
\label{density}
\end{equation}
BEC occurs when chemical potential reaches the minimum of the lower HF branch. To first order in the interactions, this condition
gives
\begin{equation}
\mu_c{=}-\hbar\Omega+C\tilde n^0(\mathbf0){+}g_{\uparrow \downarrow}\tilde n_{\uparrow \downarrow}^0(\mathbf{0}).
\label{critical_chemical_potential}
\end{equation}
where $\tilde n^0(\mathbf0)$ and $n_{\uparrow \downarrow}^0(\mathbf{0})$ are  corresponding densities at $\mathbf{r=0}$.
The interaction-induced shift of the critical temperature $\delta T_c=T_c^{\rm HF}(\Omega)-T_c(\Omega)$ can be obtained by expanding Eq.~(\ref{density}) around  $\mu=-\hbar\Omega$ and $T=T_c(\Omega)$.
This yields

\begin{align}
\delta T_c
=
\frac{
-(g+\frac{g_{\uparrow\downarrow}}{2})\mathcal{I}_1  -g_{\uparrow\downarrow}\mathcal{I}_2}{
\displaystyle
\int d^3\mathbf r
\left.\left(
\frac{\partial\tilde n_{+}^0(\bf{r})}{\partial T}
+
\frac{\partial\tilde n_{-}^0(\bf{r})}{\partial T}
\right)\right|_{\mu=-\hbar\Omega, T=T_c(\Omega)}
},
\label{fraction}
\end{align}

with 
\begin{align}
 \mathcal{I}_1 =&   \displaystyle\int d^3\mathbf r
\Bigg[\left\{\left(\tilde n_{+}^0({\bf 0})+\tilde n_{-}^0({\bf 0})\right)
-
\left(\tilde n_{+}^0(\mathbf r)+\tilde n_{-}^0(\mathbf r)\right)
\right\}\nonumber\\
\times&\left(
\frac{\partial\tilde n_{+}^0(\bf{r})}{\partial\mu}
+
\frac{\partial\tilde n_{-}^0(\bf{r})}{\partial\mu}
\right)\Bigg]_{\mu=-\hbar\Omega, T=T_c(\Omega)},
\end{align}
and 
\begin{align}
  \mathcal{I}_2 &=  \int d^3\mathbf{r}\left[
\frac{\partial\tilde n_{-}^0(\bf{r})}{\partial\mu}
\{\tilde n_{\uparrow\downarrow}^0({\bf 0})-\tilde n_{\uparrow\downarrow}^0(\mathbf{r})\}\right.\nonumber\\
&+\left.\frac{\partial\tilde n_{+}^0(\bf{r})}{\partial\mu}
\{\tilde n_{\uparrow\downarrow}^0({\bf 0})+\tilde n_{\uparrow\downarrow}^0(\mathbf{r})\}\right]_{\mu=-\hbar\Omega, T=T_c(\Omega)}.
\end{align}

Evaluating the above expression analytically (see Appendix~\ref{correctionTc} for details), we obtain the interaction-induced shift of the BE condensation temperature,

 \begin{widetext}
     \begin{equation}
         \frac{\delta T_c}{T_c(0)} = \frac{2^{1/3}N^{1/6}}{a_{\text{HO}}\sqrt{\pi}[\zeta(3)]^{1/6}}\left(\frac{T_c(\Omega)}{T_c(0)}\right)^{3/2}\frac{-(a+\frac{a_{\uparrow\downarrow}}{2})A- \frac{a_{\uparrow\downarrow}}{2}B} {\left[3\left\{g_3(e^{-2\beta\hbar\Omega}) + \zeta(3)\right\}+ 2\beta\hbar\Omega g_2(e^{-2\beta\hbar\Omega})\right]},
         \label{Tc_int}
     \end{equation}
 \end{widetext}
where 
\begin{eqnarray}
A &=&\left[\zeta(3/2) + g_{3/2}(e^{-2\beta\hbar\Omega})\right]\left[\zeta(2) +g_{2}(e^{-2\beta\hbar\Omega})\right]\nonumber\\
&-&S(\beta\hbar\Omega),\nonumber\\
B &=&\left[\zeta(3/2) - g_{3/2}(e^{-2\beta\hbar\Omega})\right]\left[\zeta(2) +g_{2}(e^{-2\beta\hbar\Omega})\right]\nonumber\\
&-&S^{'}(\beta\hbar\Omega),\nonumber\\
S(\beta\hbar\Omega) &=& \sum_{l,j=1}^{\infty} \frac{1 + e^{-2j\beta\hbar\Omega} + e^{-2l\beta\hbar\Omega}+e^{-2(l+j)\beta\hbar\Omega}}{l^{1/2}j^{3/2}(l+j)^{3/2}},\nonumber\\
S^{'}(\beta\hbar\Omega) &=& \sum_{l,j=1}^{\infty} \frac{ e^{-2j\beta\hbar\Omega} + e^{-2l\beta\hbar\Omega}-e^{-2(l+j)\beta\hbar\Omega}-1}{l^{1/2}j^{3/2}(l+j)^{3/2}},
\end{eqnarray}
with $a_{\uparrow\uparrow}=a_{\downarrow\downarrow}=a$ and $a_{\text{HO}} = (\hbar/m\omega)^{1/2}$ being the harmonic oscillator length. 

At $\Omega=0$ and $a_{\uparrow\downarrow}=0$, the system consists of two
independent, equally populated components rather than one component containing
all $N$ atoms. Equation~(\ref{Tc_int}) consequently reduces to
\begin{equation}
\frac{\delta T_c}{T_c^{0}}
=-1.326\frac{a}{a_{\rm HO}}\left(\frac{N}{2}\right)^{1/6},
\end{equation}
which is the standard trapped-gas HF result applied to each
component~\cite{georgini_1996}. Here, $T_c^0 = \hbar \omega N^{1/3}/k_B[\zeta(3)]^{1/3}$ is the critical temperature of a ideal single-component Bose gas in 3D.

\section{Numerical results}
\label{numerical}
To obtain a comprehensive picture of the temperature-driven phase transitions in a coherently coupled BEC, we numerically investigate the system using HFB theory within the Popov approximation under the semiclassical approximation. This treatment captures the evolution across the full sequence of phases, from the ferromagnetic condensate to the paramagnetic condensate and, ultimately, to the thermal cloud. In particular, we focus on the transition from the paramagnetic condensate to the thermal cloud and compare the corresponding numerical critical temperatures with the analytical normal state HF results obtained from Eq.~(\ref{Tc_int}) in the presence of interactions. 

\subsection{Self-consistent semiclassical HFB-Popov calculation}

To determine the condensate fraction for $T<T_c$ as a function of temperature, we employ the following self-consistent iterative procedure:

\begin{enumerate}
\item For a fixed total number of atoms $N$, we solve the coupled GP Eqs.~(\ref{order1}) by imaginary-time propagation to obtain the condensate wave functions $\phi_i$ and the chemical potential $\mu$.

\item Using the converged condensate wave functions and chemical potential, we construct the BdG matrix in Eq.~(\ref{bdg_matrix}) At each phase-space point $(\mathbf p,\mathbf r)$ and determine the quasiparticle excitation spectrum from Eq.~(\ref{eigen_value}). The thermal density and thermal coherence are then evaluated from Eq.~(\ref{ntilde_bdg})-(\ref{ntilde12_bdg}) using the excitation energies and the corresponding spectral weights. The total number of thermal atoms is obtained as
$N_T=\sum_i\int d^3\mathbf{r}\,\tilde{n}_i(\mathbf{r}).$

\item The condensate population is subsequently updated according to the number conservation condition $N_c = N - N_T$,
and the condensate wave functions are recalculated for the updated condensate number.

\item Steps (1)--(3) are repeated until convergence is achieved in local thermal densities.
\end{enumerate}
We consider a coherently coupled Bose gas with $g_{\uparrow\downarrow}=1.6g$ confined in a spherically symmetric harmonic trap. 
Exploiting the spherical symmetry of the system, the three-dimensional GP and BdG equations can be reduced to a set of effectively one-dimensional radial equations.
 The condensate ground state is first obtained by solving Eqs.~(\ref{order1}) in imaginary time using the Crank--Nicolson scheme~\cite{muruganandham_2008,arko_cpc_2020,kaur_2021, paramjeet_2022}. The corresponding excitation spectrum and spectral weights are then evaluated from the BdG equations on an optimally chosen momentum $k$-grid, which are subsequently used to calculate the thermal densities at different temperatures. Finally, the condensate fraction is determined through the self-consistent procedure described above. The thermal densities are updated at successive iteration of self-consistent calculations using successive under relaxation technique  to accelerate the convergence. The self-consistent iterations are terminated when the relative change in the local thermal density between two successive iterations falls below a convergence tolerance of \(10^{-3}\).
\subsection{Phase diagram and condensation temperature}
\begin{figure}[H]
    \centering
\includegraphics[width=\columnwidth]{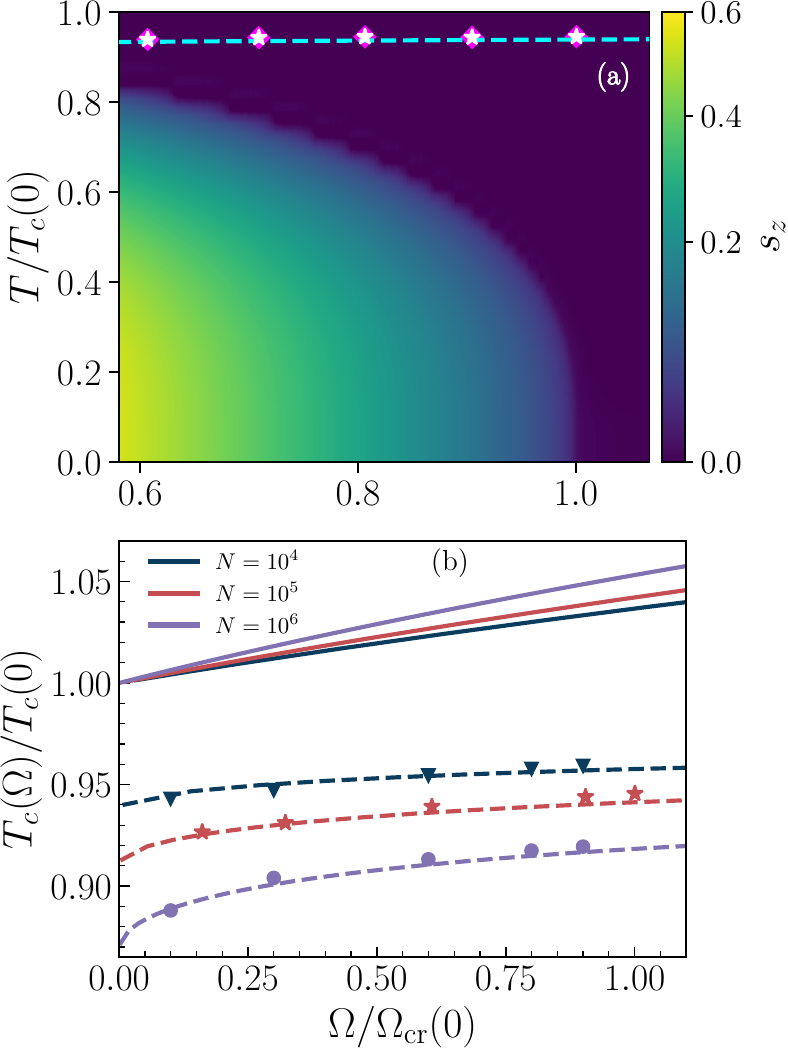}
    \caption{(a) Phase diagram of a coherently coupled BEC at finite temperature in the $T-\Omega$ plane, color coded by the magnetization $s_z$ for $N=10^5$. The vanishing of $s_z$ identifies the finite-temperature ferro--paramagnetic phase boundary. The paramagnet--thermal gas transition is indicated by the dashed line, obtained analytically from Eq.~(\ref{Tc_int}), with the corresponding critical temperatures compared with the numerical results (markers). This boundary also marks the vanishing of the transverse magnetization $s_x$. White markers denote the HFB-Popov results for the critical temperature obtained from the vanishing condensate fraction, while magenta markers indicate the corresponding temperatures at which $s_x$ vanishes, (b) Comparison of the critical temperature as a function of $\Omega$ for the interacting (dashed lines, Eq.~(\ref{Tc_int})) and ideal (solid lines, Eq.~(\ref{tc_3d})) systems for different atom numbers. The markers are the numerical HFB-Popov results. For each $N$, the temperature is normalized by its corresponding zero-coupling critical temperature, while $\Omega$ is normalized by the corresponding $\Omega_{\rm cr} = n({\bf 0})(g_{\uparrow\downarrow}-g)/2$.}
    \label{phase_diagram}
\end{figure}
At the outset, we benchmark our numerical approach by comparing the temperature dependence of the condensate fraction obtained from the HFB--Popov calculations with the corresponding analytical non-interacting result given by Eq.~(\ref{frac_3d}). To this end, we set the interaction strengths to zero and find excellent agreement for all atom numbers considered, thereby validating our numerical approach. 
We then introduce interactions, allowing us to characterize the successive transitions from the ferromagnetic phase to the paramagnetic phase and subsequently to the thermal cloud. In Fig.~\ref{phase_diagram}(a), we show the transition from the trapped ferromagnetic to the paramagnetic phase in the $\Omega$-$T$ plane, identified by the vanishing of the longitudinal magnetization $s_z = \int d^3{\bf{r}}\langle {\hat \Psi}^\dagger \hat{\sigma}_z {\hat \Psi}\rangle=0$. We additionally identify the transition from the paramagnetic phase to the thermal cloud by the simultaneous vanishing of the condensate fraction and the transverse magnetization, $s_x = \int d^3{\bf{r}}\langle {\hat \Psi}^\dagger \hat{\sigma}_x {\hat \Psi}\rangle$. Here ${\hat \Psi} = [{\hat \Psi_{\uparrow}}, \, {\hat\Psi_{\downarrow}}]^{T}$ with $\hat{\sigma}_{x,y,z}$ being the Pauli spin matrices. The numerically obtained critical line for this transition shows excellent agreement with the 
analytically predicted critical temperature given by Eq.~(\ref{Tc_int}). We emphasize that the HFB--Popov framework employed here is applicable only in the presence of a finite condensate fraction. Accordingly, the numerical results (markers) for the paramagnetic-to-thermal-cloud boundary are obtained by extrapolating the results from condensate fraction of approximately 0.3 to vanishing condensate fraction to estimate the critical temperature, as shown in Fig.~\ref{phase_diagram}(a). The numerical condensation fraction is fitted using a second order polynomial yielding the critical temperature with an estimated uncertainty of approximately $1\%$. It is worth noting that $s_y=\int d^3{\bf r}\,\langle\hat{\Psi}^\dagger\hat{\sigma}_y\hat{\Psi}\rangle=0$ throughout the entire $\Omega$-$T$ plane.

Fig.~\ref{phase_diagram}(b) illustrates the competing effects of coherent coupling and interactions on the condensation temperature. For the non-interacting system (solid lines), $T_c$ increases with the coherent coupling strength for all atom numbers considered. In contrast, interactions (dashed lines) substantially suppress $T_c$, with the suppression becoming more pronounced for larger atom numbers. Nevertheless, increasing the coherent coupling partially compensates this interaction-induced reduction, leading to a gradual increase of $T_c$ with $\Omega$ also in the interacting system. Thus, coherent coupling and interactions act in opposite directions on the condensation temperature, with the interaction-induced suppression remaining dominant over the coupling range considered.

\section{Conclusions}\label{conclusions}

We have investigated the BE condensation and finite-temperature
magnetic phase transitions in a coherently coupled Bose mixture confined
in a three-dimensional harmonic trap. For the ideal gas, we obtain
analytical expressions for the critical temperature and condensate
fraction within the semiclassical approximation. We find that coherent coupling enhances the condensation temperature for finite particle numbers, with the effect weakening as the particle number increases. For experimentally relevant atom numbers, this enhancement remains appreciable over experimentally accessible coupling strengths.
We further account for the finite-size correction to the critical
temperature. The correction systematically lowers $T_c$ relative to the
semiclassical result and decreases with increasing particle number.

For the interacting system, treated within the HFB--Popov framework, we obtain the finite-temperature phase diagram in the
temperature--Rabi coupling plane. The system exhibits successive
ferromagnetic, paramagnetic, and thermal-gas regimes. The
ferromagnetic--paramagnetic transition is identified by the vanishing of
the longitudinal magnetization, while the transition to the
thermal gas is associated with the disappearance of the condensate and
the transverse magnetization. The critical temperatures extracted
from the vanishing condensate fraction show excellent agreement with
the analytical predictions, establishing the validity of the
semiclassical HF description for locating the condensation transition.
In contrast to the ideal-gas case, the interacting system exhibits a substantial suppression of the critical temperature, with only a weak dependence on the coherent coupling strength; the suppression becomes increasingly pronounced for larger atom numbers.
An interesting direction for future work would be to provide an analytical estimate of the critical temperature associated with the onset of ferromagnetic order in the thermal cloud, which may provide a route towards an analytical description of the ferromagnetic--paramagnetic transition.

\begin{acknowledgments}
We thank Subhadeep Patra and Sivasankar PM  for
several insightful discussions. A.R. acknowledges the support of the Science and Engineering Research Board (SERB),
Department of Science and Technology, Government of India, under the project
SRG/2022/000057 and IIT Mandi seed-grant funds under the project IITM/SG/AR/87.
A.R. acknowledges the National Supercomputing Mission (NSM) for providing
computing resources of PARAM Himalaya at IIT Mandi, which is implemented by
C-DAC and supported by the Ministry of Electronics and Information Technology
(MeitY) and Department of Science and Technology (DST), Government of India.
S.G. acknowledges support from the
Science and Engineering Research Board, Department of Science and Technology, Government of India, through Project
No. CRG/2021/002597.
\end{acknowledgments}
\bibliographystyle{apsrev4-2}
\bibliography{references}

\clearpage
\onecolumngrid
\appendix

\section{DOS in 3D}
\label{dos3d}

The density of states (DOS) is defined as, 
\begin{equation}
    \rho_{3D}(E) = \sum_{\pm}\sum_{n_x,n_y, n_z}\delta(E-E_n^{\pm}).
\end{equation}
In the semiclassical approximation, the sums over the harmonic-oscillator quantum numbers can be replaced by integrals, giving 
\begin{equation}
\rho_{3D}(E)=\sum_{\pm}\iiint dn_x\,dn_y\,dn_z\,\delta(E-E_n^{\pm}).
\end{equation}\\

It is convenient to first determine the total number of states with energy below $E$, which we denote by $G(E)$. This can be written in terms of the Heaviside step function as

\begin{eqnarray}
    G(E) =  \sum_{\alpha}\sum_{n} \Theta(E-E_n^\alpha),
     = \sum_{j=+1,-1}\sum_{n_xn_yn_z} \Theta[E- \left\{\hbar(n_x \omega_x + n_y\omega_y + n_z\omega_z)+ j  \Delta\right\}] \nonumber 
    \end{eqnarray}
Under the semiclassical approximation, this expression becomes
    \begin{eqnarray}
    G(E) = \sum_{j=+1,-1}\int_{n_xn_yn_z}dn_xdn_ydn_z\Theta [E- \{\hbar(n_x \omega_x + n_y\omega_y + n_z\omega_z) {+j} \Delta\}].
   \end{eqnarray}
To evaluate the above integral, we introduce the variables
    \begin{eqnarray}
    \varepsilon_x = n_x\hbar\omega_x, \varepsilon_y = n_y\hbar\omega_y, \varepsilon_z = n_z\hbar\omega_z\nonumber,
    \end{eqnarray}
    so that the integration measure transforms according to
    $$dn_xdn_ydn_z = \frac{d\varepsilon_xd\varepsilon_yd\varepsilon_z}{\hbar^3\omega_x\omega_y\omega_z},$$

\text{and hence}, number of states becomes,
    \begin{eqnarray}
    G(E) = \frac{1}{\hbar^3\omega_x\omega_y\omega_z}\sum_j\int_{\varepsilon_x\varepsilon_y\varepsilon_z\ge0}\Theta\left[E-\left(\sum_i\varepsilon_i+{j} \Delta\right)\right]d\varepsilon_xd\varepsilon_yd\varepsilon_z,       \hspace{3cm}(i=x,y,z)\nonumber
    \end{eqnarray}
Introducing the shifted energy
\begin{eqnarray}
    \hspace{2.7cm}E^{'} = E\pm\Delta,\nonumber
\end{eqnarray}
the above expression can be written as
    \begin{eqnarray}
    \hspace{2.7cm}G(E)= \frac{1}{\hbar^3\omega_x\omega_y\omega_z}{\sum_j}\int_{\varepsilon_x\varepsilon_y\varepsilon_z\ge0}\Theta\left(E^{'}-{\sum_i}\varepsilon_i\right)d\varepsilon_xd\varepsilon_yd\varepsilon_z,  \hspace{2cm}(E^{'} = E+j\Delta)\nonumber
    \end{eqnarray}
The Heaviside function restricts the integration to the region in which the sum of the three positive energies does not exceed $E^{'}$. Thus,
\begin{eqnarray}
   G(E) = \frac{1}{\hbar^3\omega_x\omega_y\omega_z}{\sum_j}\Theta(E^{'})\int_{\varepsilon_x\varepsilon_y\varepsilon_z\ge0}^{\varepsilon_x+\varepsilon_y+\varepsilon_z\le E^{'}}d\varepsilon_xd\varepsilon_yd\varepsilon_z.\nonumber  
    \end{eqnarray}

    On solving the above integral one obtains, 
    \begin{eqnarray}
    \int_{\varepsilon_x\varepsilon_y\varepsilon_z\ge0}^{\varepsilon_x+\varepsilon_y+\varepsilon_z\le E^{'}}d\varepsilon_xd\varepsilon_yd\varepsilon_z = \int_{0}^{E^{'}}d\varepsilon_x\int_{0}^{E^{'}-\varepsilon_x}d\varepsilon_y\int_{0}^{E^{'}-\varepsilon_x-\varepsilon_y}d\varepsilon_z = \frac{(E^{'})^3}{6} ={\frac{(E+j\Delta)^3}{6}}
    \end{eqnarray}

   which gives, $$G(E) = {\sum_j}\frac{\Theta(E+j\Delta)}{6\hbar^3\omega_x\omega_y\omega_z}(E+j\Delta)^3.$$
Finally, differentiating the total number of states with respect to energy yields the density of states,
$$\rho_{3D}(E) = \frac{dG(E)}{dE} ={\sum_j} \frac{\Theta(E+j\Delta)}{2\hbar^3\omega_x\omega_y\omega_z}(E+j\Delta)^2,$$

\section{Finite size correction to 1D}
\label{fs1d}
It is well known that BEC does not occur in a homogeneous one-dimensional Bose gas in the thermodynamic limit ~\cite{Pethick_Smith_2001}. However, as shown by Ketterle and van Druten~\cite{ketterle_1996}, a finite number of atoms confined in a harmonic trap exhibits a finite-size crossover to BEC, allowing one to define a characteristic critical temperature. Motivated by this, we extend our finite-size analysis to determine the critical temperature of a non-interacting coherently coupled Bose gas confined in a one-dimensional harmonic trap.

The density of states (DOS) can be obtained by following the same procedure employed for the two- and three-dimensional cases. The single-particle spectrum of the coherently coupled system is given by
\begin{equation}
E_n^{\pm}=\left(n_x+\frac{1}{2}\right)\hbar\omega_x\pm\Delta,
\end{equation}
where $\Delta=\hbar\Omega$ denotes the coherent-coupling-induced energy splitting. The corresponding density of states is
\begin{equation}
\rho_{1\mathrm{D}}(E)
=
\sum_{\pm}\sum_{n_x}\delta(E-E_n^{\pm}).
\end{equation}
Within the semiclassical approximation, the summation over quantum numbers may be replaced by an integral,
\begin{equation}
\rho_{1\mathrm{D}}(E)
=
\sum_{\pm}
\int dn_x\,\delta(E-E_n^{\pm}),
\end{equation}
which yields
\begin{equation}
\rho_{1\mathrm{D}}(E)
=
\frac{1}{\hbar\omega_x}\Theta(E+\Delta)
+
\frac{1}{\hbar\omega_x}\Theta(E-\Delta).
\end{equation}
Thus, unlike in two and three dimensions, the one-dimensional density of states is independent of energy and depends only on the trapping frequency.
Following Ref.~\cite{ketterle_1996}, we incorporate finite-size effects by retaining the finite ground-state energy, $E_0=\hbar\omega_x/2$, as the lower limit of the energy integration. Consequently, Eq.~(\ref{exc}) becomes
\begin{eqnarray}
N&=&
\frac{1}{\hbar\omega_x}
\bigg[
\int_{\hbar\omega_x/2}^{\,2\Delta+\hbar\omega_x/2}
\frac{d\alpha}{e^{\alpha/k_BT_c}-1}\nonumber\\
&+&
\int_{2\Delta+\hbar\omega_x/2}^{\infty}
\frac{2\,d\alpha}{e^{\alpha/k_BT_c}-1}
\bigg].
\end{eqnarray}

Introducing the dimensionless variable $x=\alpha/k_BT_c$, the first integral evaluates to
\begin{equation}
k_BT_c
\left[
-\ln\left(1-e^{-\hbar\omega_x/2k_BT_c}\right)
\right],
\end{equation}
while the second yields
\begin{equation}
k_BT_c
\left[
-\ln\left(1-e^{-(2\Delta+\hbar\omega_x/2)/k_BT_c}\right)
\right].
\end{equation}
For $k_BT_c\gg\hbar\omega_x$, the first logarithm can be expanded to leading order. Consequently, we obtain
\begin{equation}
N=
\frac{k_BT_c}{\hbar\omega_x}
\left[
\ln\left(\frac{2k_BT_c}{\hbar\omega_x}\right)
-
\ln\left(1-e^{-(2\Delta+\hbar\omega_x/2)/k_BT_c}\right)
\right].
\end{equation}
The above transcendental equation can be solved numerically to determine the critical temperature for a given atom number $N$, trapping frequency $\omega_x$, and coherent coupling strength $\Omega$. In the absence of coherent coupling ($\Omega=0$), it reduces to
\begin{equation}
N=
\frac{2k_BT_c(0)}{\hbar\omega_x}
\ln\left(\frac{2k_BT_c(0)}{\hbar\omega_x}\right),
\end{equation}
which is the well-known finite-size result for a non-interacting two-component Bose gas in a one-dimensional harmonic trap~\cite{ketterle_1996}. 

\section{Correction to $T_c$ due to interactions}
\label{correctionTc}
We start from Eq.~(\ref{fraction}), which needs to be evaluated to determine the interaction-induced correction to the critical temperature. The correction can be written as

\begin{align}
\delta T_c
=
\frac{
-(g+\frac{g_{\uparrow\downarrow}}{2})\mathcal{I}_1  -g_{\uparrow\downarrow}\mathcal{I}_2}{
\displaystyle
\int d^3\mathbf r
\left(
\frac{\partial\tilde n_{+}^0(\bf{r})}{\partial T}
+
\frac{\partial\tilde n_{-}^0(\bf{r})}{\partial T}
\right)
},
\label{num_den}
\end{align}
where $\tilde{n}^0_{\pm}({\bf r})$ are given in Eq.~(\ref{n_T_non-int}) and the integrands in numerator and denominator are to be evaluated at
        $\mu=-\hbar\Omega$ and $\beta=(k_BT_c(\Omega))^{-1}$. We first consider the contribution from the first integral of numerator which is,

\begin{align}
 \mathcal{I}_1 =   \displaystyle\int d^3\mathbf r
\left(
\frac{\partial\tilde n_{+}^0(\bf{r})}{\partial\mu}
+
\frac{\partial\tilde n_{-}^0(\bf{r})}{\partial\mu}
\right)
\left[
\left\{\tilde n_{+}^0({\bf 0})+\tilde n_{-}^0({\bf 0})\right\}
-
\left\{\tilde n_{+}^0(\mathbf r)+\tilde n_{-}^0(\mathbf r)\right\}
\right].
\end{align}
As a representative term, we consider the contribution involving the central density of the two non-condensate branches. Using the expression for the thermal density and its derivative with respect to the chemical potential, we obtain

\begin{eqnarray}
    \mathrm{I.}\quad\int d^3{\bf{r}} \frac{\partial \tilde{n}_{+}^0(\bf{r})}{\partial \mu}[(\tilde{n}_{+}^0({\bf 0}) + \tilde{n}_{-}^0({\bf 0}))] &=
    &\frac{\beta}{\lambda_T^6}\left\{g_{3/2}[z_{+}^0({\bf 0})] + g_{3/2}[z_{-}^0({\bf  0})]\right\}\int d^3 {\bf{r}} g_{1/2}[z_{+}^0({\bf r})],
    \label{equation1}
\end{eqnarray}

where $z_{\pm}^0({\bf r}) = \exp[-\beta(V - \mu \pm \hbar\Omega)]$ with $\mu = -\hbar \Omega$ and employing the identity

\begin{equation}
\frac{\partial g_s(z)}{\partial z} = \frac{1}{z}g_{s-1}(z),
\end{equation}
we get

\begin{eqnarray}
\frac{\partial \tilde{n}_{+}^0(\bf r)}{\partial \mu} &=
\frac{\beta} {\lambda_T^{3}}g_{1/2}[z_{+}^0(\bf r)]. 
\end{eqnarray}

To proceed further, it is convenient to express ${\bf r}$-dependent Bose functions in terms of their standard series representations. In particular, 

$$g_{1/2}[z_{+}^0({\bf r})] = \sum_{l=1}^{\infty} \frac{e^{-l\beta(V({\bf{r}}) +2 \hbar\Omega)}}{l^{1/2}}.$$
 Using these series representations,  Eq.~(\ref{equation1}) becomes 
\begin{eqnarray}
\quad\int d^3{\bf{r}} \frac{\partial \tilde{n}_{+}^0(\bf{r})}{\partial \mu}[\tilde{n}_{+}^0({\bf 0}) + \tilde{n}_{-}^0({\bf 0})]&=&\frac{\beta}{\lambda_T^6}\left\{g_{3/2}[z_{+}^0({\bf 0})] + g_{3/2}[z_{-}^0(\bf 0)]\right\} \sum_l \frac{e^{-2l\beta\hbar\Omega}}{l^{1/2}}\int d^3 {\bf{r}} \exp[-l\beta V({\bf r})],\nonumber\\
&=& \frac{\beta}{\lambda_T^6}\left\{g_{3/2}[z_{+}^0({\bf 0})] + g_{3/2}[z_{-}^0({\bf 0})]\right\} \sum_l \frac{e^{-2l\beta\hbar\Omega}}{l^{1/2}} \left(\frac{2\pi}{ml\beta \omega^2}\right)^{3/2},\nonumber\\
&=& \frac{\beta}{\lambda_T^6}\left(\frac{2\pi}{m\beta \omega^2}\right)^{3/2}\left\{g_{3/2}[z_{+}^0({\bf 0})] + g_{3/2}[z_{-}^0({\bf 0})]\right\} g_2[z_+^0({\bf 0})],
\end{eqnarray}

where we have used the identity
\begin{equation}
\int d^3{\bf{r}} e^{-l\beta m\omega^2{\bf{r}}^2/2} = \left(\frac{2\pi}{ml\beta \omega^2}\right)^{3/2}.\nonumber
\end{equation}
The remaining contributions to the  ${\mathcal I}_1$ are as follows:
\begin{eqnarray}
&\mathrm{II.}&\quad \int d^3{\bf{r}} \frac{\partial \tilde{n}_{+}^0(\bf{r})}{\partial \mu}[\tilde{n}_{+}^0({\bf{r}}) + \tilde{n}_{-}^0({\bf{r}})] = \frac{\beta} {\lambda_T^{6}}\left(\frac{2\pi}{m\beta \omega^2}\right)^{3/2}\left[\sum_{l,j}^\infty \frac{e^{-2\beta\hbar\Omega(l+j)}}{l^{1/2} j^{3/2}(l+j)^{3/2}} + \sum_{l,j}^{\infty}\frac{e^{-2l\beta\hbar\Omega}}{l^{1/2}j^{3/2}(l+j)^{3/2}} \right],\nonumber\\
&\mathrm{III.}&\quad \int d^3{\bf{r}} \frac{\partial \tilde{n}_{-}^0(\bf{r})}{\partial \mu}[\tilde{n}_{+}^0({\bf 0}) + \tilde{n}_{-}^0({\bf 0})] =\frac{\beta} {\lambda_T^{6}}\left(\frac{2\pi}{m\beta \omega^2}\right)^{3/2}\left[\sum_{l,j}^\infty \frac{e^{-2j\beta\hbar\Omega}}{l^2j^{3/2}} + \sum_{l}^{\infty}\frac{1}{l^{2}}\zeta(3/2) \right],\nonumber\\
&\mathrm{IV.}&\quad\int d^3 {\bf{r}}\frac{\partial \tilde{n}_{-}^0(\bf{r})}{\partial \mu}[\tilde{n}_{+}^0({\bf{r}}) + \tilde{n}_{-}^0({\bf{r}})] = \frac{\beta} {\lambda_T^{6}}\left(\frac{2\pi}{m\beta \omega^2}\right)^{3/2}\left[\sum_{l,j}^\infty \frac{e^{-2j\beta\hbar\Omega} +1}{l^{1/2} j^{3/2}(l+j)^{3/2}}\right]. \nonumber\\
\end{eqnarray}

Combining the contributions from $(\mathrm I)$–$(\mathrm {IV})$, the first integral $\mathcal{I}_1$ of the numerator becomes, 

\begin{eqnarray}
 {\mathcal I}_1&=&\frac{\beta}{\lambda_T^{6}}\left(\frac{2\pi}{m\beta \omega^2}\right)^{3/2} \Bigg(\sum_{l,j}^\infty \frac{e^{-2\beta\hbar\Omega(l+j)}}{l^2 j^{3/2}} + \sum_{l}^{\infty}\frac{e^{-2l\beta\hbar\Omega}}{l^2}g_{3/2}(1) - \sum_{l,j}^{\infty}\frac{e^{-2\beta\hbar\Omega(l+j)}}{l^{1/2} j^{3/2}(l+j)^{3/2}} - 2\sum_{l,j}^{\infty}\frac{e^{-2l\beta\hbar\Omega}}{l^{1/2}j^{3/2}(l+j)^{3/2}}\nonumber \\
 &+&\sum_{l,j}^\infty \frac{e^{-2j\beta\hbar\Omega}}{l^2j^{3/2}} + \zeta(2)\zeta(3/2) - \sum_{l,j}^{\infty}\frac{1}{l^{1/2}j^{3/2}(l+j)^{3/2}}\Bigg)\nonumber\\
    &=&\frac{\beta} {\lambda_T^{6}}\left(\frac{2\pi}{m\beta \omega^2}\right)^{3/2}\left[ \left\{\zeta(3/2) + g_{3/2}(e^{-2\beta\hbar\Omega})\right\}\{\zeta(2) + g_{2}(e^{-2\beta\hbar\Omega})\} - S(\beta\hbar\Omega)\right]\nonumber\\
    &=&\frac{\beta} {\lambda_T^{6}}\left(\frac{2\pi}{m\beta \omega^2}\right)^{3/2}A,
    \label{first_integral}
\end{eqnarray}

where
\begin{eqnarray}
A =\left[ \{\zeta(3/2) + g_{3/2}(e^{-2\beta\hbar\Omega})\}\{\zeta(2) + g_{2}(e^{-2\beta\hbar\Omega})\} - S(\beta\hbar\Omega)\right],\nonumber\\
S(\beta\hbar\Omega) =  \sum_{l,j}^{\infty}\frac{1 + e^{-2j\beta\hbar\Omega} + e^{-2l\beta\hbar\Omega} + e^{-2(l+j)\beta\hbar\Omega}}{l^{1/2}j^{3/2}(l+j)^{3/2}}.
\end{eqnarray}
\\
Similarly, 

\begin{eqnarray}
  \mathcal{I}_2 &=&\int d^3\mathbf{r}\left\{
\frac{\partial\tilde n_{-}^0(\bf{r})}{\partial\mu}
[\tilde n_{\uparrow\downarrow}^0({\bf 0})-\tilde n_{\uparrow\downarrow}^0(\mathbf{r})]
+\frac{\partial\tilde n_{+}^0(\bf{r})}{\partial\mu}
[\tilde n_{\uparrow\downarrow}^0({\bf 0})+\tilde n_{\uparrow\downarrow}^0(\mathbf{r})]\right\}\nonumber\\
 &=&\frac{\beta}{ \lambda_T^{6}}\left(\frac{2\pi}{m\beta \omega^2}\right)^{3/2}B,
 \label{second_integral}
\end{eqnarray}
where
\begin{eqnarray}
B=[ \{\zeta(3/2) - g_{3/2}(e^{-2\beta\hbar\Omega})\}\{\zeta(2) + g_{2}(e^{-2\beta\hbar\Omega})\} - S^{'}(\beta\hbar\Omega)], \nonumber\\
S^{'}(\beta\hbar\Omega) = \sum_{l,j=1}^{\infty} \frac{ e^{-2j\beta\hbar\Omega} + e^{-2l\beta\hbar\Omega}-e^{-2(l+j)\beta\hbar\Omega}-1}{l^{1/2}j^{3/2}(l+j)^{3/2}}.
\end{eqnarray}

We now evaluate the denominator of Eq.~(\ref{num_den})  at $\mu=-\hbar\Omega$ and $T=T_c(\Omega)$ via

\begin{equation}
\int d^3 {\bf{r}}\left[\frac{\partial \tilde{n}_{+}^0(\bf{r})}{\partial T} +\frac{\partial \tilde{n}_{-}^0(\bf{r})}{\partial T} \right]_{\mu=-\hbar\Omega, T=T_c(\Omega)}.
\end{equation}
 The temperature derivative of the noncondensate density corresponding to the $+$ branch is given by,

\begin{eqnarray}
\frac{\partial \tilde{n}_{+}^0(\bf{r})}{\partial T} &=& \frac{\partial}{\partial T} \frac{g_{3/2}[z_+^0({\bf r})]}{\lambda_T^3}\nonumber\\
&=&\int d {\bf{r}}\left[ \frac{3}{2 T}\frac{g_{3/2}[z_+^0({\bf r})]}{\lambda_T^3}  +  \frac{\beta(V({\bf{r}})+2\hbar\Omega)}{T\lambda_T^3}g_{1/2}[z_+^0({\bf r})]\right ],\nonumber\\
&=& \frac{3}{2T \lambda_T^3}\sum_j\frac{e^{-2\beta j\hbar\Omega}}{j^{3/2}}\int d{\bf r} e^{-j\beta V({\bf r})} +\frac{\beta}{T\lambda_T^3}\sum_j\frac{e^{-2\beta j\hbar\Omega}}{j^{1/2}} \int d{\bf r} V({\bf r}) e^{-j\beta V ( {\bf r})}+\frac{2\beta\hbar\Omega}{T\lambda_T^3}\sum_j\frac{e^{-2j\beta\hbar\Omega}}{j^{1/2}}\int d{\bf r}e^{-j\beta V({\bf r})}\nonumber,\\
&=&\frac{1}{T\lambda_T^3}\left(\frac{2\pi}{\beta m \omega^2} \right)^{3/2}\{3 g_3[z_+^0({\bf 0})]+2\beta\hbar\Omega g_2[z_+^0({\bf 0})]\}.
\label{first_denom}
\end{eqnarray}

Similarly
\begin{equation}
\int d^3{\bf{r}} \frac{\partial \tilde{n}_{-}^0(\bf{r})}{\partial T} = \frac{3}{T \lambda_T^3}\left(\frac{2\pi}{\beta m \omega^2} \right)^{3/2}\zeta(3).
\label{second_denom}
\end{equation}

Adding Eq.~(\ref{first_denom}) and Eq.~(\ref{second_denom}), we get
\begin{eqnarray}
 \int d^3{\bf{r}} \left[\frac{\partial \tilde{n}_{+}^0(\bf{r})}{\partial T} +\frac{\partial \tilde{n}_{-}^0(\bf{r})}{\partial T} \right] = \frac{1}{\lambda_T^{3}T}\left(\frac{2\pi}{\beta m \omega^2} \right)^{3/2}\left[3\left\{g_3(e^{-2\beta\hbar\Omega}) + \zeta(3)\right\}+ 2\beta\hbar\Omega g_2(e^{-2\beta\hbar\Omega})\right].
 \label{denomenator}
 \end{eqnarray}

Finally, using Eqs. (\ref{first_integral}), (\ref{second_integral}), and (\ref{denomenator}) in Eq.(\ref{num_den}), we get,

\begin{align}
\delta T_c = \frac{-[(g+\frac{g_{\uparrow\downarrow}}{2})A+\frac{g_{\uparrow\downarrow}}{2}B]}
{{\lambda_T^{3}}{k_B}\left[3\left\{g_3(e^{-2\beta\hbar\Omega}) + \zeta(3)\right\}+ 2\beta\hbar\Omega g_2(e^{-2\beta\hbar\Omega})\right]}.
\end{align}
The above expression can be further simplified by expressing the interaction strengths in terms of the corresponding scattering lengths, $g=4\pi\hbar^2 a/m$ and $g_{\uparrow\downarrow}=4\pi\hbar^2 a_{\uparrow\downarrow}/m$, and normalizing the temperature correction by the critical temperature of the two-component Bose gas, $T_c(0)=\hbar\omega (N/2)^{1/3}/[k_B\{\zeta(3)\}^{1/3}]$. This yields the interaction-induced correction to the critical temperature of the coherently coupled BEC

     \begin{equation}
         \frac{\delta T_c}{T_c(0)} = \frac{2^{1/3}N^{1/6}}{a_{\text{HO}}\sqrt{\pi}[\zeta(3)]^{1/6}}\left(\frac{T_c(\Omega)}{T_c(0)}\right)^{3/2}\frac{-[(a+\frac{a_{\uparrow\downarrow}}{2})A+ \frac{a_{\uparrow\downarrow}}{2}B]} {\left[3\{g_3(e^{-2\beta\hbar\Omega}) + \zeta(3)\}+ 2\beta\hbar\Omega g_2(e^{-2\beta\hbar\Omega})\right]}.
         \label{append_Tc_int}
     \end{equation}

\end{document}